Technical Report

# *Compact Low-Profile Wearable Antennas For Breast Cancer Detection*

**GROUP MEMBERS:**

Hamad AlShehhi        100036762

Mariam Alzarouni        100038629

Noura AlYammahi        100038625

**PROJECT SUPERVISOR:**

Prof. Raed Shubair

Dr. Nazar Ali

**SUBMISSION DATE:**

## I. Acknowledgement

This project gave us the opportunity to explore a new field and apply our communication and electronic engineering knowledge in the medical field to help in saving people's lives. In addition, it helped us in enhancing our communication skills, through the group members weekly meetings, the discussion with the supervisors, the presentations and poster that we presented, and via email. Also, in this project we were able to perform tasks beyond the limit of our expertise.

The reasons behind the success of this project are the cooperative, flexible, independent and responsible team members. In addition to the great project supervisors Professor Raed Shubair and Dr. Nazar Ali, who were very supportive and motivate supervisors, as they considered themselves as a part of this project. They inspired us with their feedback and helpful comments.

In addition, we are grateful to Khalifa University of Science, Technology and Research for the extraordinary support whether it was by providing us with everything the project required or the encouragement to do our best. As a team, we have gained a huge experience by working on this project and we have gained a good idea about engineering in real life.

## Table of Contents

# List of Tables and Figures

## List of Tables:

## List of Figures:

## II. Abstract

As the number of breast cancer patients are increasing, we need more reliable method that can helps in detecting this type of cancer in early stages, which will lead to save patient's lives as the earlier the tumor is detected the higher the chance for recovery. Therefore, the aim of our project is to design a Compact Low-Profile Wearable Antenna that is used for early breast cancer detection with low cost.

In this project, we will use a micro-strip patch antenna which is the most suitable type of antenna for this project due to its small size, light weight, low profile, easy fabrication and low cost. Furthermore, the frequency band is the ISM band, which is the Industrial, Scientific and Medical band is used in this project (2.4GHz to 2.5GHz). Moreover, the High Frequency Structure Simulator (HFSS) software is used in our project. The reason behind choosing this software is its high accuracy and speed, high performance for computation, automatic solution process, and powerful post-processing capabilities.

# CHAPTER ONE

# **INTRODUCTION**

## 1.1. MOTIVATION

Many lives can be saved if tumours are detected in early stages, which can result in a bigger chance for recovery. Many patients find it irritating to get regular check ups due to the fact that the majority of the monitoring systems are complicated, not available everywhere and not mobile. Further more, for medical field applications, micro-strip antennas are efficient and have flexible properties that are utilized in imaging, diagnosis and treatment.

It is known that breast cancer is the most common type of cancer in the world, and the earlier its been detected the better. In the early stages of breast cancer, getting rid of from the tumours is much easier and more guaranteed. Nowadays, the main method that is used in the hospitals for breast cancer detection is the Ultra Wideband method (UWB). However, Many patients find it irritating to get regular check ups due to the fact that the majority of the monitoring systems are complicated and not mobile.

## 1.2. AIM

The aim of our project is to Develop Compact Low-Profile Wearable Antennas to help us to detect the breast cancer earlier with low cost, fast, and use the screen to show the tumor.

## 1.3. OBJECTIVES

1- How the antenna will help to detect the breast cancer.

2- Investigation of existing the antennas strategy for MICS and ISM bands.

3- Study a various alternatives for breast cancer detection using microwave imaging.

4- Using the experiences in order to design a prototype for UWB compact micro strip patch antenna array.

5- After getting an obtained data for real human tissues. We are able to test the efficiency of this result in the lab.

# CHAPTER TWO

# COMMON ANTENNAS FOR HEALTHCARE APPLICATIONS

## 2.1. OVERVIEW

An antenna is a reciprocal device that transmits and/or receives electromagnetic waves and signals; also it can be considered as transducer device as it converts the radiated electromagnetic waves to electrical signals, and via versa. In order to convey the electromagnetic energy from the source to the transmitting antenna, or from the receiving antenna to the target, waveguide is needed, which is the transmission line. Figure 1 below illustrates the path of the electromagnetic waves from the source to the transmitting antenna through the transmission line, and it is noticeable that the antenna joins between wire (transmission line) and wireless (free-space).

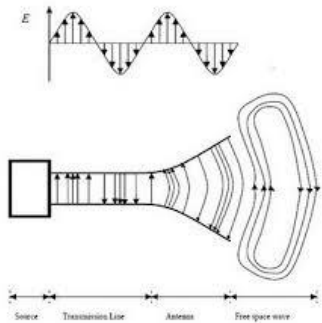

**Figure 1:** Transmitting Antenna [1]

In the medical field generally and cancer specifically, antennas have considerable implementations, as it is used for information transmission, diagnosis and treatment. For instance, wireless capsule endoscopy has been daily used in the hospitals for information transmission. Moreover, the Radio Frequency Identification (RFID) is used in most of the implantable monitoring systems. Additively, there are many other medical applications that antennas are used in such hyperthermia treatments, medical implants, microwave imaging, and wireless wellness monitoring. Therefore, antennas can be classified as a communication device and/or a treatment device.

Recently, antenna's research major purpose is to minimize the size and reduce the complexity of the antenna design.

The study that is base on the connection between diversified electronic devices in or on the human body is called The Wireless Body Area Networks (WBANs) study. The WBAN consists of a group of tiny nodes that are equipped with biomedical sensors, motion detectors, and wireless communication devices which incorporates techniques similar to those implemented in wireless systems [1-12]. It is also known that at the microwave frequency band, the dielectric constant and conductivity of the human body became high. Furthermore, for the purpose of safeguarding the human body from being affected by the radio waves, the Specific Absorption Rate (SAR) is used in the wireless devices. Therefore, the WBAN requires flexible antennas with compact size, low impact on the human body and small SAR. For instance, patch antennas are used for the on or off body communication, due to their small profile, high gain and simple design. On the other hand, the patch antenna has an obstacle, which is the huge back radiation [13-24].

## 2.2. INTRODUCTION

Globally it is known that Breast cancer is the main prevalent type of cancer in the world, and the earlier it's been detected the better. Presently, the most common way for the detection is X-ray imaging, which is called the X-ray mammography. However, due to the limits in the X-ray mammography, researchers started to search for a new advanced device that will aid in the early detection of the Breast cancer. For long time ago, microwaves were utilized for breast cancer detection, and the microwave imaging functions by using low power electromagnetic wave at microwave frequencies in order to illuminate the target, and then view its internal parts. Utilizing microwaves radiation into the breast in order to check and discover the existence of tumors, is a popular method and it can be divided into two approaches the tomography and the radar-based. In the first stage, tomography, one transmitter is used to radiate toward the breast, in the same time, around the breast several antennas are laid in order to receive scattered and diffracted waves. Based on the received data, some processes are done to get a 2-D or 3-D image of the breast. The radar-based microwave imaging is another used method for breast cancer detection. This method requires a single ultra-wideband (UWB) antenna is used to transmit a short pulse to the breast and receive the scattered signal back, and the same thing is repeated in various locations around the breast. If the strength of the scattering was high, it will indicate that there is a tumor. In this method extra detailed information are provided compared to the tomography microwave imaging method. However, the UWB antennas should authorize resolution level, which causes some limitations. Thus, the major characteristics that must be taken into consideration are large fractional bandwidth, low side-lobes and low mutual coupling. In addition, to minimize the reflection between the free space and the breast surface, the metallic sensors must be placed in a matching medium that has almost the same permittivity of the breast tissue. However, in real life, the implementation of the matching medium around the patient's breast is hard [25].

## 2.3. MICRO-STRIP PATCH ANTENNA

Micro-strip antennas recently are used in the medical field for the purpose of imaging, diagnosis, and treatment. In addition, the flexibility in the micro-strip antenna leads to the ability of contacting between the human body and the skin.

There are three main parts in the micro-strip antenna, as it is shown in Figure 2 below.

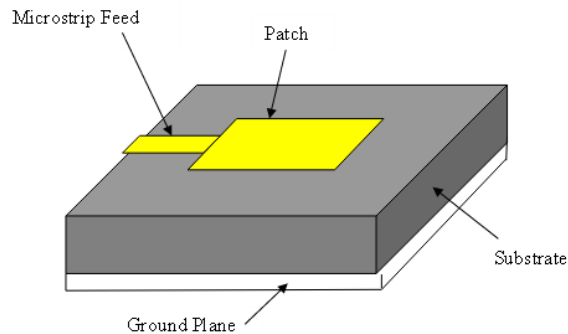

**Figure 2:** Micro-Strip Patch Antenna [25]

Firstly, the ground plane which consists of a conductor. Secondly, the substrate part which includes the dielectric, and finally, a conductor that represents the metal patch. In the design of the antenna, the first two components (the ground plane and the substrate) are over each other and they both have the same exact size. However, the metal patch component has almost half the size of the other two components [25].

Antennas need power to work, and this power is provided to the antenna through a copper wire called the feeder. The feeder is linked to the first part of the antenna (the metal patch), which is conductor, thus current will flow and voltage will appear. Reaching to the middle point of the antenna, the voltage will approach zero, while at the end points the current will have zero value. Having a very small metal patch, will lead to having an open circuit, and thus the current will be zero [26].

### 2.3.1 MICRO-STRIP PATCH ANTENNA IN BREAST CANCER

The early detection of breast cancer can be done via the microwave imaging. Due to the micro-strip patch antenna's features such as small size, light weight, low profile, and low manufacturing cost, several studies are done on the capability of micro-strip patch antenna in detecting the breast tumor. Therefore, micro-strip patch antenna became active and useful for the imaging, as it will aid in the early detection of the cancer and it will help to locate the suspicious lesion in the breast for a biopsy procedure [27].

### 2.4. IMPLANTABLE ANTENNA

For the purpose of communication, inductive coupling is used in the implanted medical devices, which causes problems long meter distances transmission. In order to over come this issue, for wide distances, the transmission can be done via a wireless device that send the information as electromagnetic signals. Therefore, in the implantable devices for the aim of communication between the patient and the access point (long distance), a small helical folded dipole antenna is utilized [28].

### 2.4.1 IMPLANTABLE ANTENNA IN BREAST CANCER

Wireless technologies are implemented in the implantable antennas, in order to allow the doctors to monitor the patient's body all the time and recognize the changes that happen to the cells "including the mutation of cells into malignant tumors". Additively, there are some external antennas and other internal ones, and the shape of the antenna relies on its location [29].

### 2.5. REMARKS

These types of antennas are the mostly used in medical applications. Thus, it is very important to understand their features and characteristics in order to be able to design them in an effective and efficient way.

# CHAPTER THREE

# DESIGN CHALLENGES OF WEARABLE ANTENNAS

## 3.1. OVERVIEW

This chapter demonstrates the challenges that will be faced while designing a wearable antenna device. Some of the most likely arising challenges in this design, including size limitations, which reduce the efficiency of the antenna, battery life and other different common challenges will be introduced.

## 3.2. INTRODUCTION

Regard to the challenges, any goal may have some limitations that must be evaluated and analyzed in order to obtain the desirable result. One of the reasons that may cause these limitations is that the human body contains considerable volume of lossy conducting materials which will lead to the following various challenges that will occur once the antenna is placed.

## 3.3. REDUCTION OF ANTENNA EFFICIENCY BECAUSE OF SIZE LIMITATIONS

Due to the relationship between the performance of an antenna and its size, some limitations may arise. Therefore, the dielectric constant must be reduced in order to obtain a high performing antenna, which will lead to an increase in the thickness of the dielectric substrate. As a result, a larger bandwidth will be needed for power radiation and an increment in the antenna's size.

## 3.4. STRONG MULTI-PATH LOSSES AND LOSSY ENVIRONMENT

As the medical application demands, the antenna must be implemented in the human body after designing it. However, because of the characteristic of human bodies, which have lossy conducting materials, it will dissipate energy, which contradicts the aim of reducing power dissipation. Moreover, the illustration of the antenna's behavior in two different cases is in figure 3 and figure 4 in the next page [29].

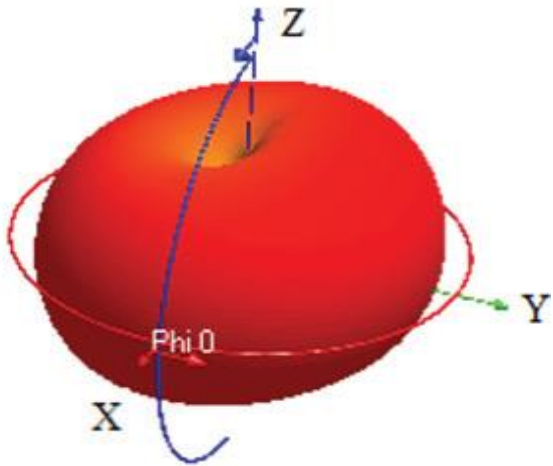

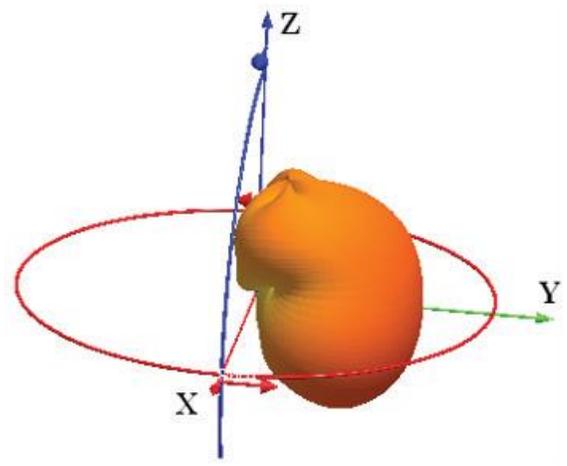

**Figure 3:** Antenna Radiation Behavior Before Implementation into a Human Body [29]

**Figure 4:** Antenna Radiation Behavior After Implementation into a Human Body [29]

It is noticeable that both figures illustrate the difference between the radiation behavior of an antenna before and after implementing it in a human body. In addition, the results of figure 3 were obtained by using the HFSS Software, while figure 4 by implanting an antenna in the skin tissue of an anatomical human head model.

### 3.5 BATTERY LIFE

Due to the high demand for significant performance through perfect quality and speed and the high usage of devices, the energy of the battery will be highly consumed. Moreover, in some wearable antennas, size of battery might create an issue, since it might limit the integration of the sensor. Therefore, in order to produce an efficient system, there will be a need for larger batteries to keep the small sensors powered.

### 3.6 CHALLENGES IN THE DESIGN OF MICROWAVE IMAGING SYSTEMS FOR BREAST CANCER DETECTION

In order to obtain better resolution, higher frequencies must be employed which will lead to lower electromagnetic field penetration within lossy tissues due to the use of small antenna elements. Moreover, the high resolution increases the size of the electromagnetic problem, which will result in an increment in the computational time needed for the image reconstruction. In addition to the previous difficulties, each imaging method also faces certain challenges which are related to the system configuration and imaging scenario, because several techniques are based on various physical effects [29].

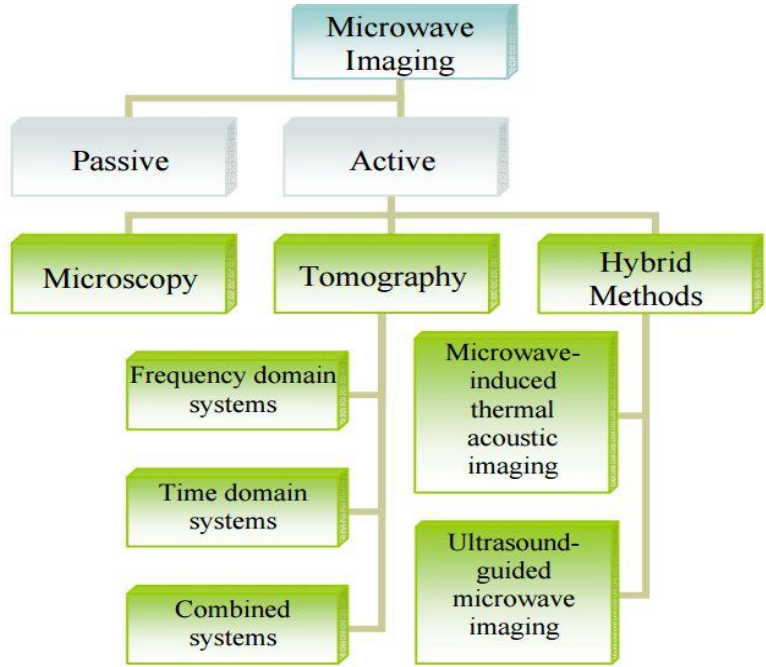

**Figure 5:** Microwave imaging systems for breast cancer diagnosis

Figure 5 above demonstrates that in passive systems, the object produces the energy. However, in active systems, the sensing can be achieved by probing the biological object with self-generated energy. Also, Active microwave imaging has several types such as microwave tomography, hybrid modalities, microwave microscopy, and ultra wideband (UWB) radar techniques [29].

## 3.7. REMARKS

Several designs and simulations for recent wearable antennas have a promising evolution in healthcare applications. These designs are being developed with significant studies, which are taken under some important consideration such as safety requirements, especially biocompatibility.

# CHAPTER FOUR

# DESIGN FEATURES OF WEARABLE ANTENNAS

### 4.1.OVERVIEW

Within this chapter the design features of wearable antennas will be discussed. First, we will start with a short introduction in order to confirm the importance of wearable antennas. As these antennas are the most significant part in the wireless transmission [29]. After that, we will discuss minimizing the physical size and reflection coefficient, maximizing the electric size, in addition to the omnidirectional radiation pattern, and finally biocompatibility.

### 4.2.INTRODUCTION                                        In

order to finish the project successfully, it is very important to recognize the design features of it. Wireless sensing systems are introduced to the medical field and applications, and the main component in it is the wearable antennas. As the measured or sensed data needs to be delivered with rapid speed in a reliable was over a wireless channel in order to analyze these data [30]. Therefore, as these antennas will interact with the human body, a careful design process that considers wearability and a highly reliable communication link must be considered [31].

### 4.3.DESIGN FEATURES

In wireless communication the important part is the transmission of the power and data from the implantable or wearable antenna to the external recording equipment, Figure 6 below shows an example of the links in the wireless system, between the implantable antenna, human tissue and an external recording device.

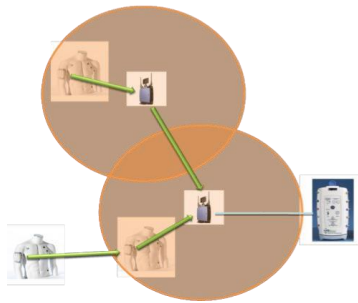

**Figure 6:** Implanted antenna communication links [30]

The most important thing to take into consideration is the patient's safety. In addition to that, the low cost, small size and reliability. Thus specific design features are considered in order to achieve that, and they are as follows:

### 4.3.1. Minimized Physical Size

The antenna must be small enough, in order to be able to integrate it in a low-profile wearable device. Large size antennas will not have a practical design; in addition to that it is not easy to design large antennas with reliable and efficient high data rate transceivers. Further more, small size antennas (millimeters) reach optimal power transfer efficiency with high frequency ranges (Gigahertz) [32]. According to that, there are three minor procedures that needs to be done, as the following:

#### *4.3.1.1. Shorting pin*

Using a shorting pin to ground the patch therefore the metal patch layer will be shorted.

#### *4.3.1.2. Slotting*

Slotting can be defined as cutting the metal patch into slots as shown in figure 7 below. So, by applying slotting the current scattering will be reduced or removed, also it will helps in keeping the current in the device for longer time. Therefore, the purpose of slotting is saving current and controlling the current path on the patch surface.

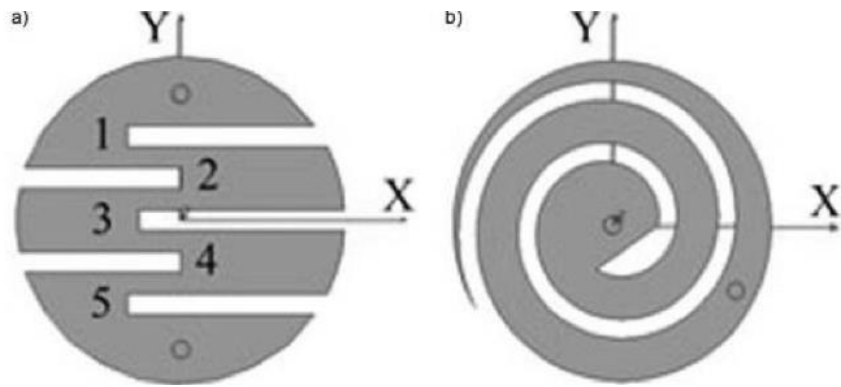

**Figure 7:** Slotting shapes on the patch surface: (a) meandered, (b) spiral [32]

### *4.3.1.3. Stacking*

In order to stack more layers, the number metal patches and substrates is increased, so the antenna will be structured as ground - substrate - metal patch - substrate - metal patch and so on. Additively, as it is shown in figure 8 below, all metal patches will be slotted each with the reverse direction of the one under it.

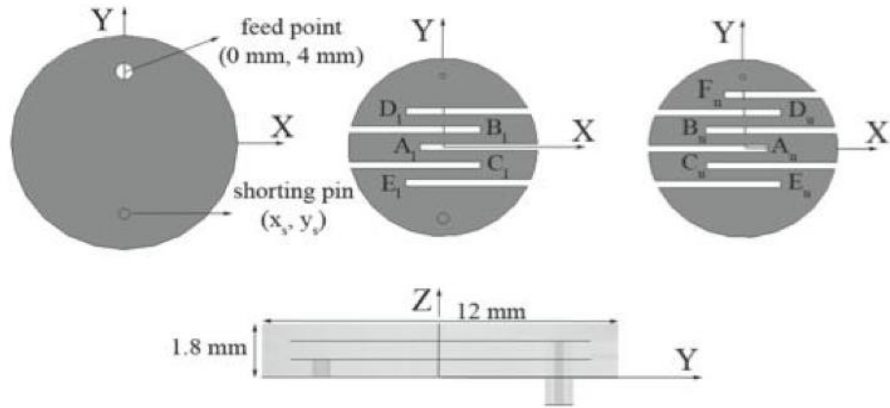

**Figure 8:** Structure of stacked wearable antenna with meandered antenna patches [32]

As it is noticeable from the above figure, by applying the stacking, the antenna will be denser, however, it is better that loosing the current by having it scattered. As a result, in order to save the current, more slots are been added. Therefore, by that increasing the current path will be guaranteed.

### 4.3.2. Minimized Reflection Coefficient

Increasing the transferred power and minimizing the power reflected from the antenna will lead to reducing the reflection coefficient. That is done by utilizing the impedance matching method, and this method consists of matching the load impedance to the source impendence (ZL = ZS) [33].

### 4.3.3. Increased Electrical Size

In healthcare and wearable devices, having a big size antenna is not favorable. However, by changing the electrical length of the antenna, we will be able to control its physical size. That is done by, placing an inductor or a capacitor in series with the antenna, so by changing the load the electrical length of the antenna wire will be controlled. This method is called lumped-impedance tuning or loading [34].

### 4.3.4. Omnidirectional Radiation Pattern

By controlling the switch state we will be able to control the antennas radiation and having an antenna that radiates in all directions. Therefore, a monopole antenna can have an omnidirectional radiation pattern or a dipole antenna with reflector as shown in figure 9 below [35].

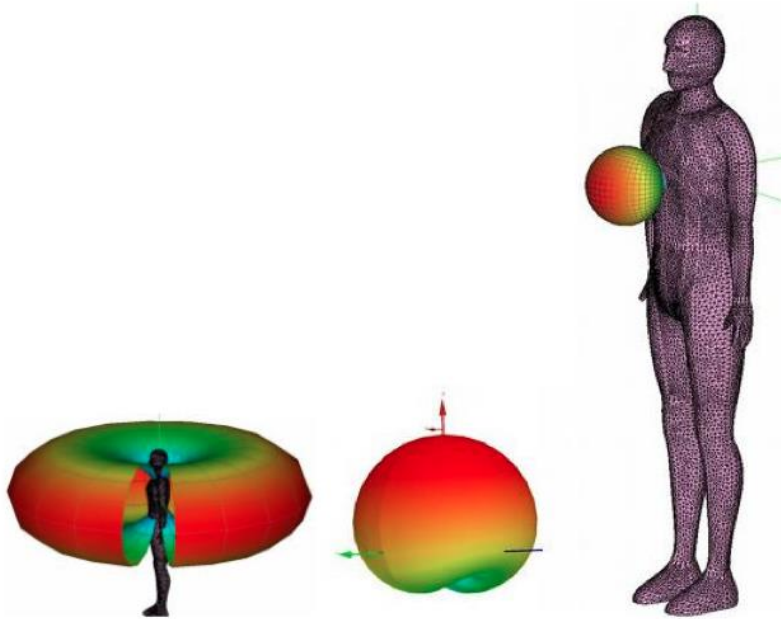

**Figure 9:** Implanted antenna with omnidirectional radiation pattern [35]

### 4.3.5. Biocompatibility

Biocompatibility is very significant feature that must be taken into consideration, because in order to ensure patient safety, the design of the antenna has to be biologically compatible in shape and size. Circular shape of the antenna is one of the best shapes as it is without any edges [36].

**4.4. REMARKS**

Providing a wireless flexible communication between the patients and the monitoring devices will provide freedom for patients. On the other hand, this will require using low-cost and low-profile antennas [37].

# CHAPTER FIVE

# MEASURING THE PERFORMANCE OF ANTENNA DESIGN

**5.1. OVERVIEW**

Due to the availability of variety of antennas types, different antenna devices are widely used. There are some essential terms that determine the performance of the antenna such as the efficiency of the antenna. In this section, the discussion of these basic characteristics of antenna will be introduced.

**5.2. INTRODUCTION**

There are different parameters that measure the performance of the antenna, including the return loss, standing wave ratio, and specific absorption rate .In addition, the performance can be influenced by the operating resonance frequencies, for example, ISM and MICS bands which are specified for healthcare applications.

**5.3 RETURN LOSS ($S_{11}$)**

$S_{11}$ is the reflection coefficient, return loss and gamma, which demonstrate the amount of power that is reflected from the antenna. When all power is reflected from the antenna (no radiation), then S11=0 dB. As an example, when S11= -10 dB, this indicates that when 3 dB power is transmitted to the antenna, the reflected power will be -7 dB, while the remaining power is transmitted to the antenna. This transmitted power may be absorbed as losses within the antenna or radiated. In addition to, the majority will be radiated because of the low loss design of the antenna. By using the Vector Network Analyzer (VNA), S11can be measured and plotted [38].

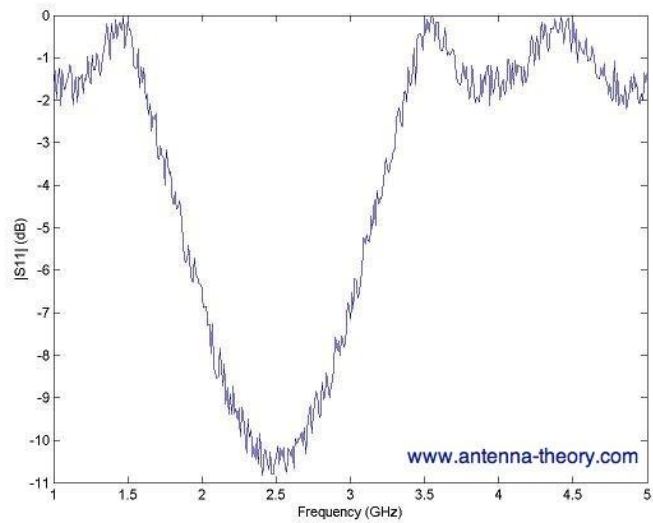

**Figure 10:** The plot of S11 versus the frequency [37]

As illustrated in Figure 10, the best antenna radiation is when S11=-10 dB at 2.5 GHz. Moreover, there will be no virtually antenna radiation at 1.5 GHz when S11 is almost 0 dB where all power is reflected. When the return loss increases, the antenna performance also increases. Finally, the maximum accepted S11= -10 dB, but the ideal value is -30 dB.

## 5.4 STANDING WAVE RATIO (SWR)

In order to deliver power from radio transmitter or receiver to the antenna, the impedance of all antenna, radio or transmission line must be similar. The numerical measurement that represents how matched the antenna's impedance with the radio or transmission line's impedance is called the standing wave radio. SWR is a function of the reflection coefficient, which illustrates the reflected power from the antenna.

The small value of the SWR means that the antenna's impedance is matched better with the radio or the transmission line impedance. In addition it also refers to, more power delivered to the antenna in the form of electromagnetic radiation. The converse is also true. It is known that the smallest value of the SWR is 1, which means that all the power is delivered to the antenna and there is no reflection and this is an ideal case. Generally, antenna match is considered good if the SWR is below 2 [39].

Standing wave ratio can be referred as VSWR (voltage standing wave ratio), and this ratio can be measured as illustrated in Figure 11 below.

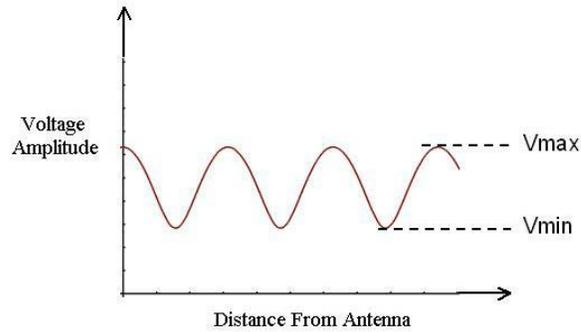

**Figure 11:** voltage Standing wave ratio measurement [39]

From the above figure, it is demonstrated that the VSWR is the ratio between the maximum amplitude of the standing wave to the minimum amplitude of the standing wave.

## 5.5 SPECIFIC ABSORPTION RATE (SAR)

SAR stands for Specific Absorption Rate, which is the measure of the amount of absorbed power by the human body, and it can be calculated using the following $SAR = \dfrac{\sigma |E|^2}{\rho}$ equation. As illustrated in the equation, it is noticeable that the SAR depends on the electrical conductivity, rms electric field and the mass density.

The dielectric property and conductivity of both the normal breast and a breast with tumor are different. For instance, according to some experiments done by applying freq. of 6 GHz, the values of the dielectric and conductivity are illustrated in the table below.

**Table 1:** dielectric and conductivity for normal breast vs breast with tumor

|  | Fat | Tumor |
|---|---|---|
| $\varepsilon_r$ | 9.5 | 46 |
| $\sigma$ (S/m) | 0.4 | 3.4 |

From Table 1, it is noticeable that both $\varepsilon_r$ (dielectric constant) and $\sigma$ (conductivity) have much higher value in the present of tumor.

Therefore, SAR will be helpful in recognizing the location of the tumor in the human body. Both the accuracy in detecting the tumor and the SAR intensity depend on the direction of antenna radiation pattern. Therefore, with a wide beam width antenna the tumor detection is better. In addition, the accuracy of the tumor location detection is increased if the radiation pattern is toward the tumor.

One of the most preferable UWB antenna type for these measurements is the bowtie patch antenna. The bowtie patch antenna is symmetric with the shape of half bow; also at the end of the patch there is ground plate, as it is shown in the figure below.

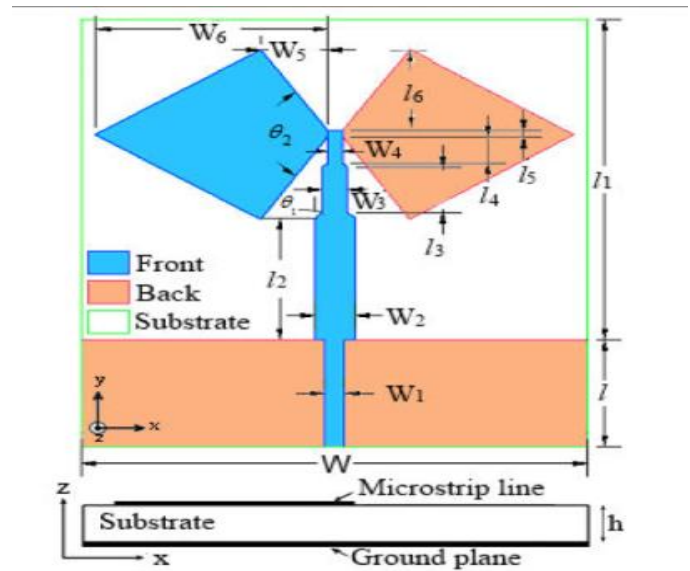

**Figure 12:** bowtie patch antenna [40]

Figure 13 below demonstrates the radiation pattern of the bowtie patch antenna in both electric (E) and magnetic (H) fields, over different range of frequencies (4-8 GHz).

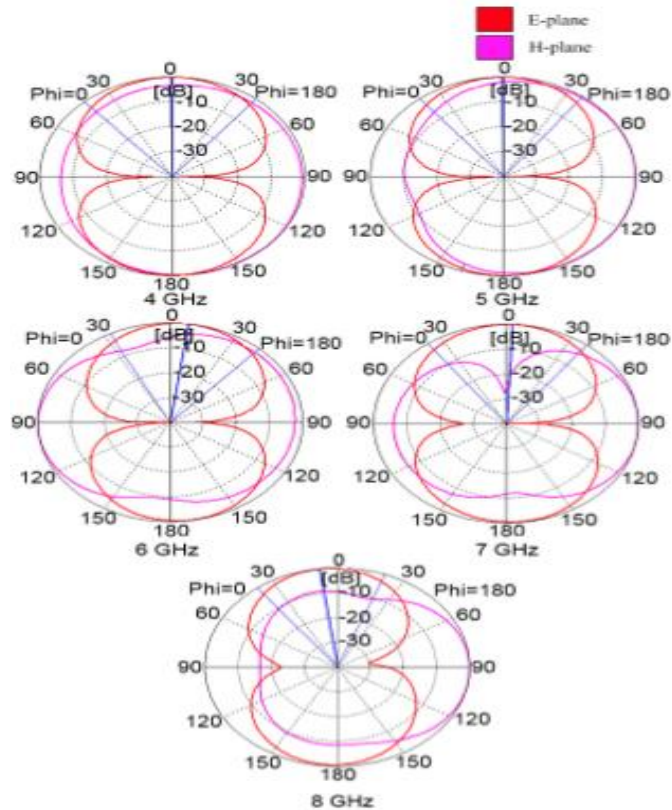

**Figure 13:** Radiation pattern for bowtie patch antenna [40]

From the above figure, considering the E-plane, it is noticeable that at 4 GHz the beam width is the widest. However, in order to choose a suitable frequency, comparison between the SAR value in breast without tumor and SAR value in breast with tumor. The maximum the difference in SAR value between breast with tumor and breast without tumor gives the best frequency that will gives accurate detection [40].

Therefore, SAR is helpful in recognizing the location of the tumor in the human body. Both the accuracy in detecting the tumor and the SAR intensity depend on the direction of antenna radiation pattern. Therefore, with a wide beam width antenna the tumor detection is better. In addition, the accuracy of the tumor location detection is increased if the radiation pattern is toward the tumor.

## 5.6. HIGH PERMITIVITY DIELECTRIC SUBSTRATE

High permittivity dielectric substrate must be used in the design in order to decrease the dimensions of the antenna. Further than, by using the high-permittivity, the effective wavelength can be shortened which will permit the resonant frequency to be shifted to a lower frequency [41].

## 5.7 OPERATING FREQUENCY: MEDICAL ISM BAND

For medical telemetry, it can be found in different bands such as: Private Land Mobile Radio Service (PLMRS), Industrial, Scientific and Medical (ISM), Wireless Medical Telemetry Service (WMTS). However, for this project the most suitable band is the ISM band.

ISM band (Industrial, Scientific and Medical band) is a part of radio spectrum that has various uses in several countries. In 1985, the FCC Rules opened the ISM bands for two main purpose, which are wireless LANs, and mobile communications. After few years, a new band with 5 GHz was added, and this additional band is called Unlicensed National Information Infrastructure (U-NII). However, the most important freq. in the ISM band is 2.4 GHZ.

ISM band consist of three different bands: 900 MHz band (33.3 cm) 2.4 GHz band (12.2 cm) 5.8 GHz band (5.2 cm). The most used spectrum is the 2.4 GHz band. For instance, it is used for microwave ovens, cordless phones, military radars, industrial heaters, and in medical telemetry. In addition, the 2.4 GHz ISM band is known as the best band for wireless applications [42].

Therefore, due to the research we did about different medical bands and due to the above reasons, we chose the ISM band for this project.

## 5.8. REMARKS

The parameters mentioned above are crucial to measure the antenna's performance which is important for the design procedure, since they ease the evaluation of the validity of the design.

# CHAPTER SIX

# USING ANTENNAS IN BREAST CANCER DETECTION

## 6.1. OVERVIEW

Chapter six will discuss existing examples and experiments about trying to use different types of antennas for breast cancer detection. This is like a literature survey or a small research we did in order to get more familiar with the topic and the medical terms and conditions.

## 6.2. INTRODUCTION

Many women in the world suffer from breast cancer. The early detection of the breast cancer increases the chances for recovery. People all around the world are researching, inventing and experimenting new early detection methods for this type of cancer. One of the famous fields is using antennas in breast cancer detection. A lot of researches and tests had been done, however there is still no 100% accurate, safe and compact device.

## 6.3. EXISTING EXAMPLES

### 6.3.1 Example 1: 3-D Antenna Array Design

The major problem while designing a fixed 3-D array antenna is the assumption that the antenna is radiating to the free space. However, for Breast Cancer detection the antenna should radiate toward the tumour in order to copy the dielectric properties of the tumour. Therefore, the design of antennas used for Breast Cancer detection must be enhanced so that it radiate directly into higher dielectric media. Further more, sampling resolution must be taken into consideration, because antennas will allow us to know the characteristics of the field distribution within the imaging volume, which will be used later in order to remodel the dielectric properties of the area it radiated in. Thus, it is desired to increase the number of antennas so that the covered surface will increase. However, increasing the number of antennas will increase the coupling between them, which is not preferable.

Additively, there are several negative affects of the mutual impedance on the performance of the antennas, such as the shift resonant frequency, alter radiation pattern and reducing the matching performance. Moreover, these affects have a different impact on each antenna in the array due to their locations. Thus, the design of the array must be done in a way to reduce the mutual coupling between the antennas and it is necessary to repeat the simulations and measurements on the prototype to enhance and develop the final design.

A tapered patch-imaging chamber was developed in the shape of the bra to make it easier for the client to use. Figure 14 below illustrates the imaging chamber and the model that was simulated using the HFSS software. This imaging chamber consists of 6-element array of patch antenna and it is placed in a lossless dielectric material [43].

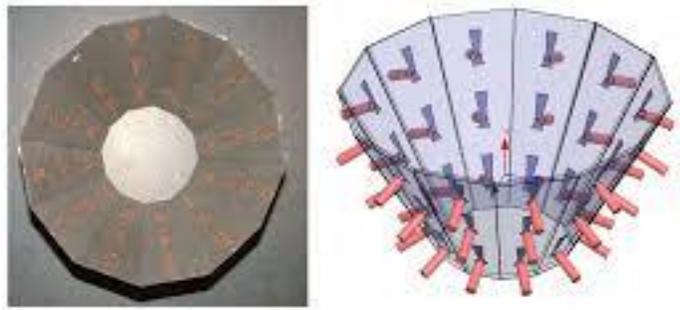

**Figure 14:** imaging chamber and the HFSS model [43]

Figure 15 below, shows the return loss for both the HFSS simulation and the measurement.

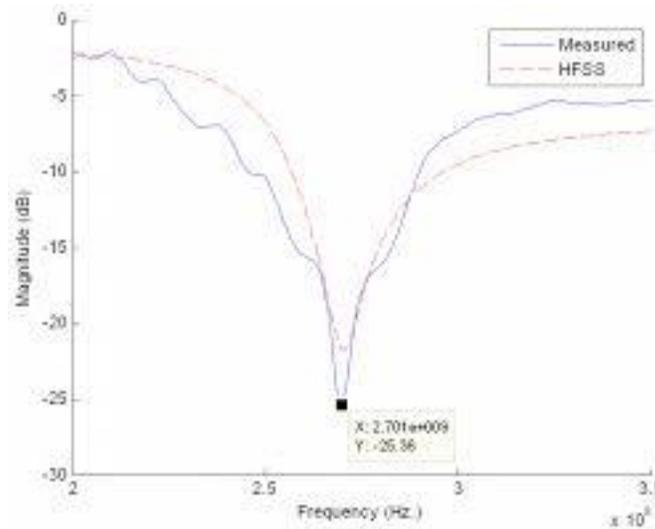

**Figure 15:** Return loss comparison [43]

It is noticeable that the HFSS simulation is similar to the real measurement for the antenna operating at a frequency of 7 GHz, which makes HFSS accurate simulation software.

A model of healthy breast and a breast with tumour were created depending on the dielectric properties of both as illustrated in Figure 16 below.

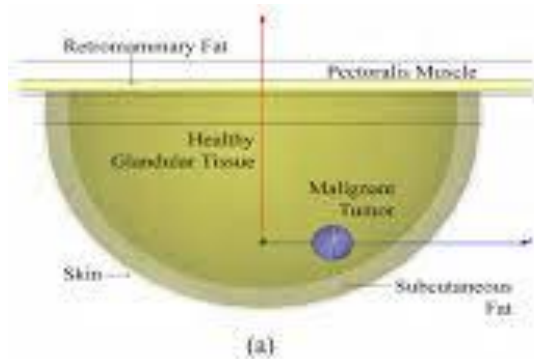

**Figure 16:** Human breast model [43]

In addition, as it is shown in Figure 17 below, the antennas are radiating effectively, and they accomplished sufficient coupling into the breast tissue. It is illustrated that the image resolution is high as the small tumors results in large scattered field [43].

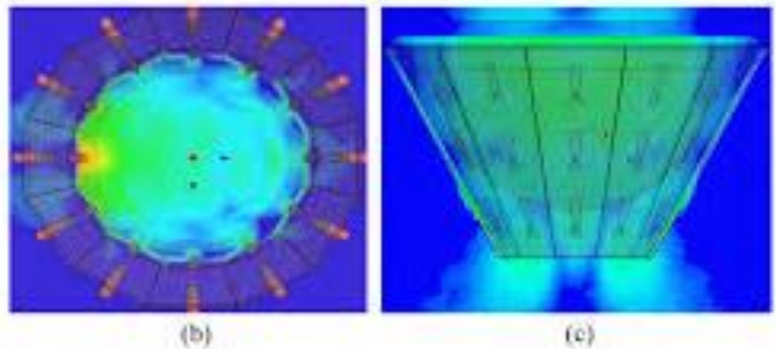

**Figure 17:** Antenna's radiation in the breast model [43]

### 6.3.2 Example 2 Symmetrical Antenna Array

The microwave radar was developed for breast cancer detection. This type of antenna array has a curved surface, which can be used for real breast cancer patients and realistic 3D breast models. The side view of its geometry with the curved breast phantom is shown below in Figure 18, while Figure 19 represents the top view of the antenna array.

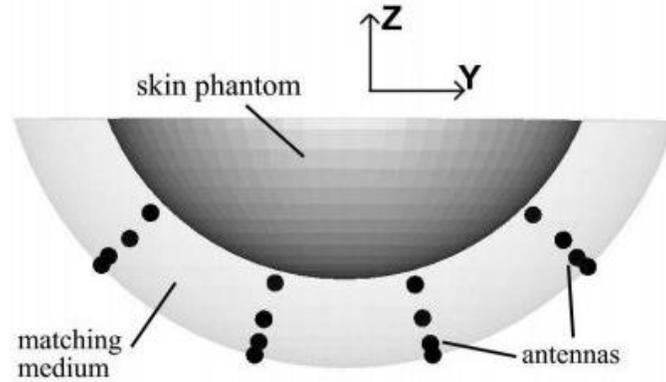

**Figure 18:** The side view for antenna array geometry with the curved spherical skin model [44]

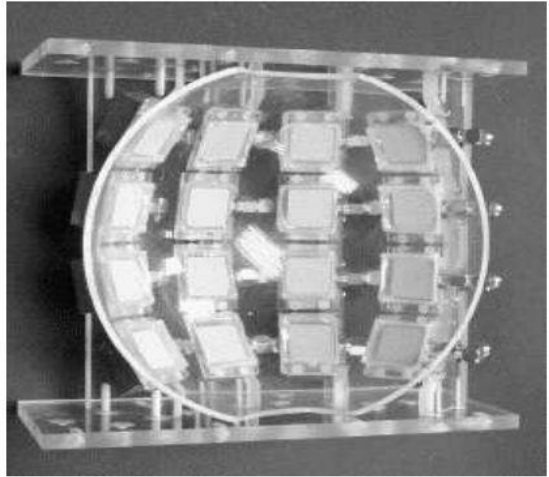

**Figure 19:** The top view of the curved spherical skin model with the antenna array [44]

As illustrated above, there are four antennas in four rows in the lower part (78 mm radius sphere). There is a clearance for the feed connectors and cables that allow symmetrical array geometry, since the antenna design is small.

The production of the realistic curved breast phantom was done, as all phantom materials have electrical properties that demonstrate realistic values of real breast. The following values are at 6GHz frequency [44].

**Table 2:** Permittivity and attenuation at freq. of 6 GHz.

|  | Breast fat/matching medium | Skin phantom | Tumour phantom |
|---|---|---|---|
| Permittivity | Close to 10 | 30 | Close to 50 |
| Attenuation | 0.8 dB/cm | 16 dB/cm | - |

During experiments, firstly, the array was filled with matching medium, while the curved phantom takes place in the right position. After that, a tank to the top of the antenna array will be attached in order to fill it with the breast fat equivalent liquid (equilivant to the matching medium).

**Radar signal processing (multi-static mode)**

The pre-processing of the raw measured radar signal is the first step of signal processing. The steps for pre-processing are as following:

1) The tumour response will be extracted from measured data.
2) Equalisation of tissue losses.
3) Equalisation of radial spread.

Post-reception synthetic focusing will be employed, in order to obtain the 3D image of the scattered energy. The focusing algorithm is relay on the sum (DAS) beam forming and classical delay.

The results for various phantom tumour detection experiments will be illustrated below in three different graphical formats [44].

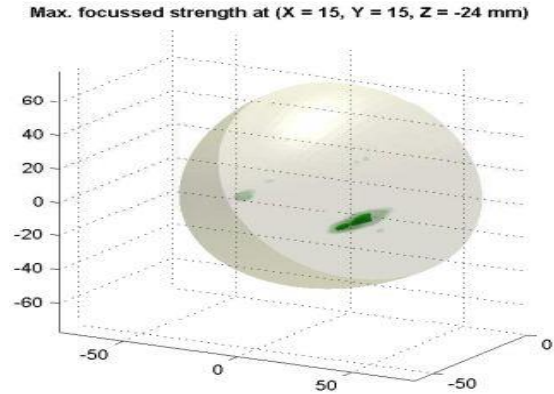

**Figure 20:** 3D scattering energy contour map (- 1.5dB isovalue threshold) [44]

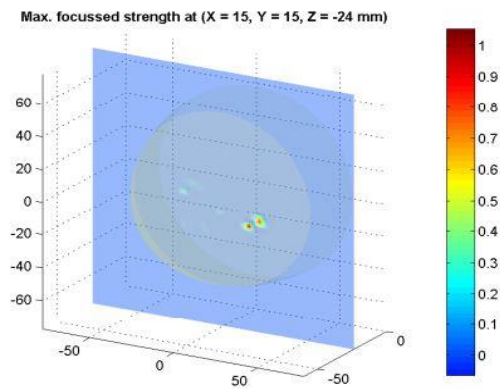

**Figure 21:** 2D planar slice through the detected tumour location [44]

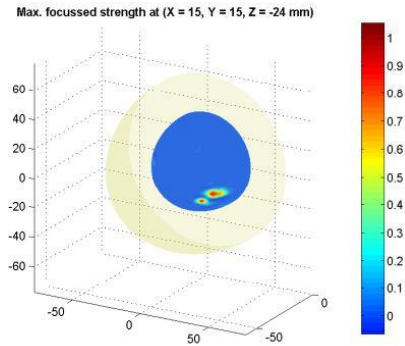

**Figure 22:** 2D radial slice through the detected tumour location [44]

As we can notice, there is clear detection of 10mm spherical tumour phantom, as it is located approximately 26mm from the skin layer.

### 6.3.3 Example 3: UWB Antennas for Near –Field Imaging

This type of antenna is a printed monopole antenna, which is fed by a 50Ω CPW (Coplanar Waveguide) that can operate in a lossy coupling medium. The geometry of this antenna is shown in Figure 23.

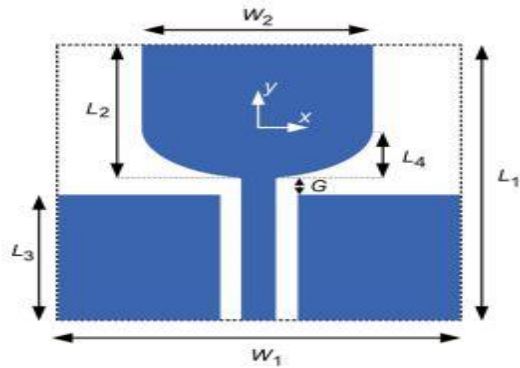

**Figure 23:** The geometry of the printed monopole antenna [45]

The rectangular patch is considered as the radiating element while the antenna's base is a semicircle that leads to a smooth transition between different resonant modes. Therefore, there will be a good impedance match within a broad range of frequency.

$$L_2 = 6.2 \times 10^7 / (f_L \times \sqrt{\varepsilon_r}) \quad [\text{m}]$$

The lowest operating frequency is $f_L$

The antenna's length is $L_2$

The relative permittivity of the medium around the antenna is $\varepsilon_r$

One of the most important problems in UWB imaging systems that confront the design of the antenna is the air-skin interface reflection. the design of the 50Ω CPW was done by using a conformal mapping method. In HFSS software, using the HFSS optimization tool to obtain the best input matching within the operating band performed the optimization of the geometrical parameters. Assuming Er 4.3, the antenna's length $L_2$ was set to 10.7mm corresponding to 2.8GHz of lowest frequency.

The simulation of the antenna in the HFSS ver. 9.2 is illustrated in Figure 24. There are two different feed models. The first one is the ideal lumped gap source 50 Ω while the second model is the subminiature version A connector (SMA).

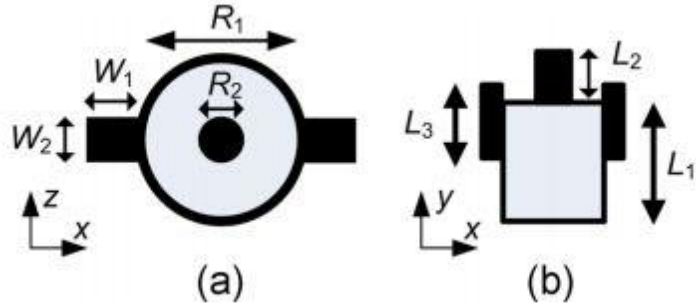

**Figure 24:** SMA connector geometry: (a) cross section in x-z plane, (b) cross section in x-y plane[45]

The return loss of the antenna for various models is simulated and measured by using HP8722ES network analyzer, as demonstrated in Figure 25. It indicates that the return loss is less than -9.6 dB (from 3.45 GHz to 9.9 GHz).

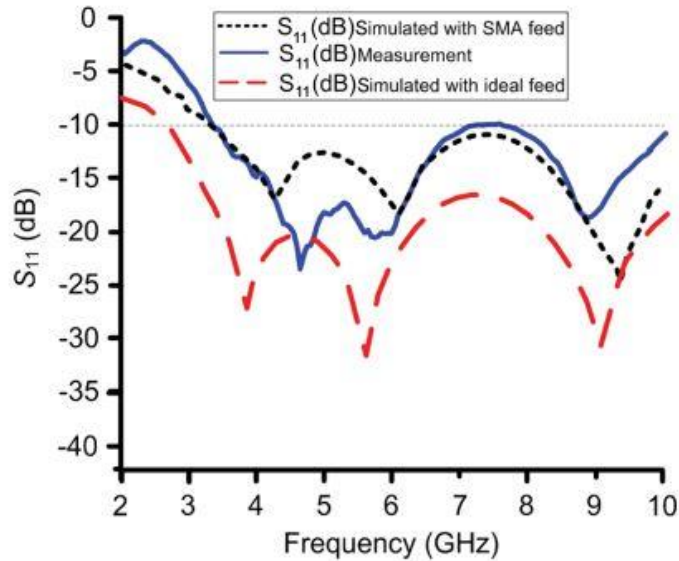

**Figure 25:** S11 for two different feed models [45]

Regard to the detection of tumor-mimicking target, an antenna array that contains two antennas in the x-y plane take place side by side. Besides placing the antenna array in the coupling medium, an Anritsu-MP1763B Pulse Pattern Generator is connected to the transmitting antenna, while an Agilent 86100A wide bandwidth oscilloscope is connected to the receiving antenna. Also, the transmitter antenna involves 200 ps Gaussian pulse and 1 V, while the receiver antenna records the backscattered signal by utilizing the wide bandwidth oscilloscope. A tumor is placed opposite to the antenna array which is mimicked with a plastic cylinder. The cylinder contains distilled water with a diameter of 0.8 cm and 1 cm length, however, the depth of the tumor varies from 0 to 6 cm. Moreover, the computation of the tumor response is done by subtracting the recorded backscattered signal when there is no tumor from the recorded backscattered signal in the presence of tumour. In addition, the system gain which is the tumor response energy normalized to the transmitted pulse energy is shown in Figure 26 below [45].

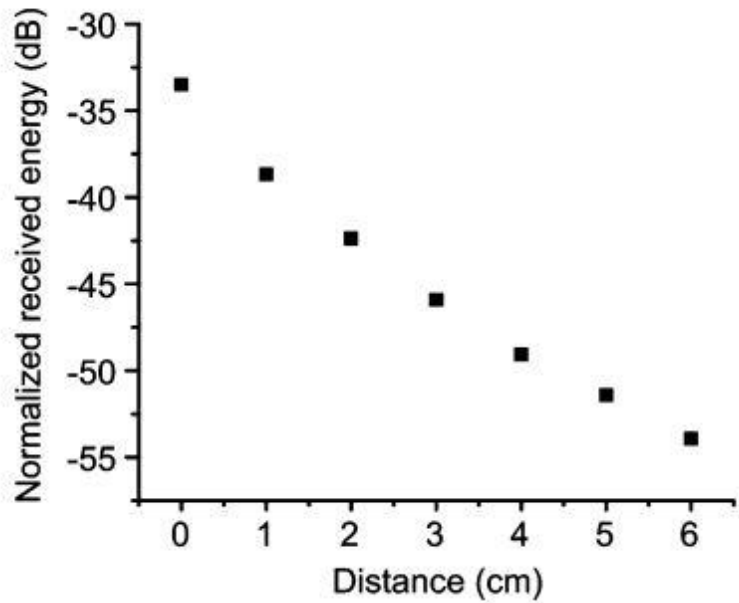

**Figure 26:** Normalized received measured energy for various tumor depths [45]

It is noticeable that, at 0 cm, the tumor response normalized energy is -33dB while it has an approximate reduction of 4 dB per cm when there is an increment in the tumor depth.

## 6.4 REMARKS:

After studying these research papers and analyze them, we got advance knowledge about the project and the how the antenna is used in breast cancer detection. In addition, this research helped us deciding the type of antenna that is most suitable for this project, the frequency band to operate on and finally the software to use.

# CHAPTER SEVEN

---

# ABOUT HFSS

## 7.1. OVERVIEW

HFSS software will be introduced in this chapter, with its special features and main advantages. In addition, the reasons behind choosing this software in this project and the usefulness of using it in antenna designing will be discussed.

## 7.2. INTRODUCTION

The software that will be used in this project is the High Frequency Structure Simulator (HFSS). It has several significant features, as it provides a high standard simulation of 3D full-wave electromagnetic fields. Furthermore, different applications can be done with this software such as microwave applications, radio frequency (RF), and integral equation for digital high-speed solutions. Also, it provides high accuracy, speed, frequency, which is required for this project [46].

## 7.3. ADVANTAGES

There are several advantages of using HFSS software, including the high speed and high accuracy. Moreover, it is useful for high frequency circuit architects, so it saves time and cash for improvement stage. Another advantage is that it has an automatic solution process, as the user is only requested to identify the property of the material and allocate the wanted outcome.

## 7.4. MAIN FEATURES

Some of the main features of the HFSS software are as follows:
• Automated Adaptive Lattice
• Solver Advances
• Advanced Finite Array Simulation Technology
• Mesh Element Technologies
• Advanced Broadband SPICE Model Generation
• Optimization and Measurable Analysis
• EDA Design Flow Incorporation
• High Execution Computing

**7.5. HFSS FOR ANTENNA DESIGN**

It is challenging to design an antenna with a small size, limited channel bandwidth, and to prevent the interaction between this antenna and other components. However, using HFSS software will save the time, enhance the results and provide us with an accurate simulation of the antenna. For instance, this software can be used in order to analyze the basic performance such as return loss, input impedance, gain, directivity and some other polarization characteristics. Furthermore, it illustrates 3-D antenna with the radiation patter, in addition to the electric and magnetic field. Modeling, simulating and analyzing antennas in an easy way that saves time make HFSS software a great choice [47].

**7.6. REMARKS**

According to the discussion above in this chapter, HFSS software is the most suitable software that will help in designing the wearable antenna required for this project.

# CHAPTER EIGHT

# BASICS OF MICRO-STRIP PATCH ANTENNA DESIGN

## 8.1. OVERVIEW

This chapter give us a background about the micro-strip patch antenna which is very important to reach the goal of the project. This chapter also, will discuss the design specification of the patch antenna.

## 8.2. INTRODUCTION

As a group, we had several meeting in order to choose the type of antenna. The challenge was to specify an antenna, which used for similar medical telemetry. Consequently, we ended with micro-strip patch antenna. The reason behind that it enables us to print directly in the circuit board. Furthermore, there are some important properties that make this type of antenna useful, such as low cost, low profile and easy to fabricated. [48]

## 8.3. DESIGN OF THE PATCH ANTENNA

As it is illustrated in the figure below, the antenna considered to two types of dielectrics. SIW layer and patch layer. The micro-strip patch antenna is located in the top of the patch layer. On the other hand, in the bottom surface there no metallization and from the metallization of the SIW layer we created the ground plane of the micro-strip patch antenna. To make the antenna simple, we use RWG instead of ISW [48].

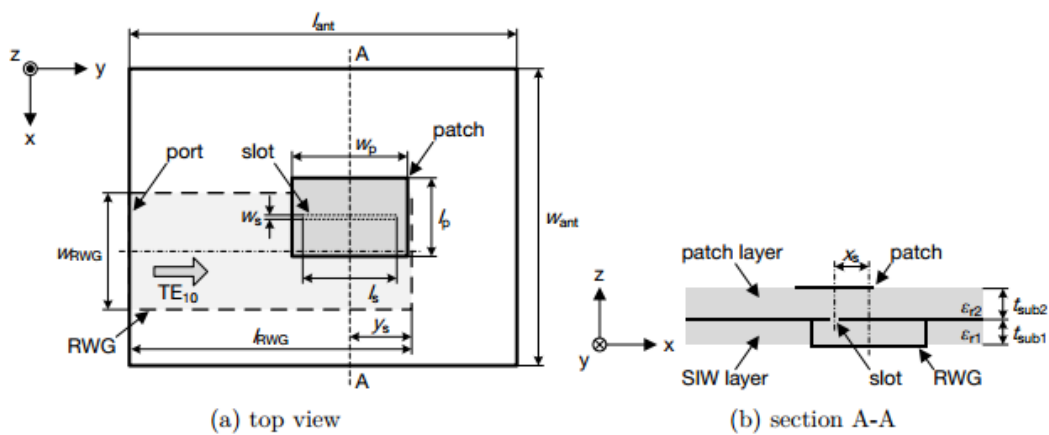

(a) top view  (b) section A-A

**Figure 27:** Structure of aperture-coupled micro-strip patch antenna fed by SIW [47]

There are three main parts in the micro-strip antenna, as it is shown in Figure 28 below.

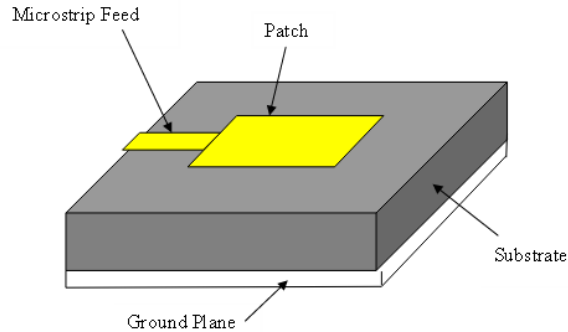

**Figure 28:** Micro-Strip Patch Antenna [49]

Firstly, the ground plane which consists of a conductor. Secondly, the substrate part which includes the dielectric, and finally, a conductor that represents the metal patch. In the design of the antenna, the first two components (the ground plane and the substrate) are over each other and they both have the same exact size. However, the metal patch component has almost half the size of the other two components [49].

Antennas need power to work, and this power is provided to the antenna through a copper wire called the feeder. The feeder is linked to the first part of the antenna (the metal patch), which is conductor, thus current will flow and voltage will appear. Reaching to the middle point of the antenna, the voltage will approach zero, while at the end points the current will have zero value. Having a very small metal patch, will lead to having an open circuit, and thus the current will be zero [50].

### 8.3.1 MICRO-STRIP PATCH ANTENNA IN BREAST CANCER

The early detection of breast cancer can be done via the microwave imaging. Due to the micro-strip patch antenna's features such as small size, light weight, low profile, and low manufacturing cost, several studies are done on the capability of micro-strip patch antenna in detecting the breast tumor. Therefore, micro-strip patch antenna became active and useful for the imaging, as it will aid in the early detection of the cancer and it will help to locate the suspicious lesion in the breast for a biopsy procedure [51].

**8.4 REMARKS**

We have to notice that, the theoretical concepts in the future maybe adjust. Because. The knowledge is improve and the people invent things better than now. On the other hand, may this project will help to find another way that help to detect the breast cancer earlier.

# CHAPTER NINE

# MULTIBAND

# ANTENNAS

## 9.1. OVERVIEW:

In this chapter, we will discuss a new type of antenna which is multiband antenna. So, the dissection will be about the multiband wearable antennas design and characteristics. Then about the dual-band wearable antenna, tri-band and quad band.

## 9.2. INTRODUCTION

There are an important effect of the multiband frequency which is help shape of the antenna and rise it is efficiency. The best description of the multiband antenna is an antenna constructed to work in many bands. These types of antennas is using design in order to active in one band. On the other hand, the other part is work in another band. Sometimes, the gain of the multiband antenna is lower than average or physically larger in compensation [52].

## 9.3. MULTIBAND ANTENNA:

## 9.3.1. MULTIBAND WEARABLE ANTENNAS DESIGN AND CHARACTERISTICS

The goals below will help to reach a simple and scalable solution to a multi-band wearable antenna: The ground plane is not eligible. This will lead to make the antenna comparatively dense and then impractical. There is only one exception, which is patch antenna, however the narrow bandwidth and pattern only in one hemisphere are not optimal.

## 9.3.2. DUAL-BAND WEARABLE ANTENNA

Dual antenna is very important in the medical field. The reason behind that is it can scan and expose many parameters at the same time. For example, the person can note the body temperature and the blood pressure together. These options will be help to save the life of people [53].

A dual-band antenna is constructed and simulated as rectangular shapes. So, it is required two rectangular shapes, which are connected with 'arms'. Many different ways used in order to implement the dual-band. One of them is rectangular method with the specific frequency and finally bridging the gap between them.

The steps below are used in order to design the dual-band using rectangular method:

**Step1**: calculate the width (W).

**Step2:** Finding the Effective dielectric constant.

**Step 3:** calculate the  Effective length ( Leff ).

**Step 4:** Calculation of Length extension (ΔL).

**Step 5:** find the value of  actual length of the patch (L).

**Step 6:** Calculation of ground plane dimensions (gL and gW) [55].

The figures below illustrates simulated and measured reflection coefficient for three types of multiband antenna, which are tri-band, dual-band and quad-band.

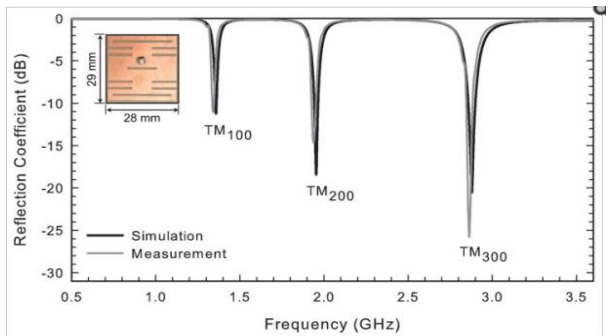
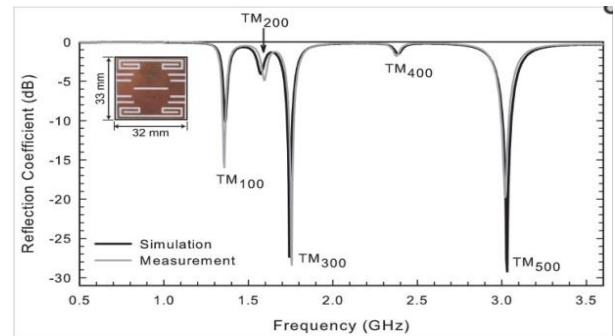

**Figure 29:** Dual-band result [55]          **Figure 30:** Tri-band result [55]

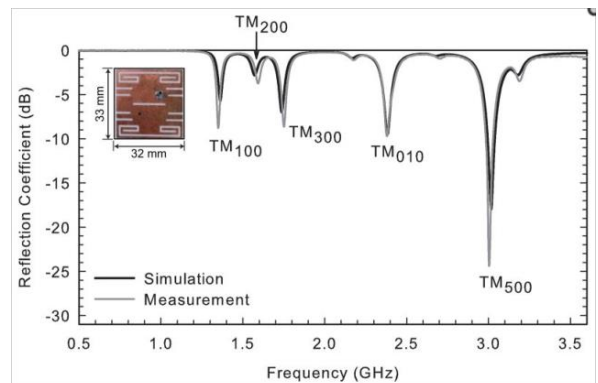

**Figure31:** Quad-band result [55]

# CHAPTER TEN

# RETURN-LOSS

# ENHANCMENT

**10.1. OVERVIEW**

In this chapter, the discussion will be about some recent technique that could enhance the return-loss, such as the meta-material structures (MTMs).

**10.2. INTRODUCTION**

There are various types of substrates that can be used in Micro-strip Patch Antennas. They are dependent on the design requirements, which depend on feed distances, length, width, relative permittivity, and the operating frequency. Therefore, some parameters such as the return loss should be taken into consideration with a view to enhance it as possible to improve the antenna performance. The return loss is defined as the power loss in the reflected signal by a discontinuity in a transmission line.

**10.3. META-MATERIALS (MTMs)**

By using meta-material structure, a significant enhancement of the return loss will be achieved which will enhance the quality of communication. Electromagnetic Meta Materials (MTMs), in recent years, had the attraction of a lot of scientists, since various research activities with remarkable achievements have been made and applied to different conventional devices and antennas in order to improve the performances [56-59]. MTMs are considered as smart materials, as they produce properties that do not occur naturally. There were developed from the composition of multiple elements from combined materials in order to have the ability of bending electromagnetic radiation around an object, while hiding the appearance, as it seems that it is not there at all. This phenomenon is called cloaking. Moreover, electromagnetic radiation is consisting of perpendicular planes of magnetic and electric fields. Usually, natural materials only impact on the electric field, but meta-materials have the ability to affect both the electric and magnetic components. This will lead to the advantage of expanding the range of interactions that are able to be done [60].

**10.4. REMARKS**

Meta-materials (MTMs) help in enhancing the return loss, which is the most significant parameter of the antenna design. Therefore, this will lead to measure the validity of the design as it verifies that the loss of power has a minimal value.

# CHAPTER ELEVEN

---

# ANTENNA DESIGNS AND RESULTS

## 11.1. OVERIVIEW

Within this chapter, all the designs we have done to reach to the final design will be introduced with their return loss graphs and 3-D polar plots.

## 11.2. INTRODUCTION

Using HFSS for designing antennas is very helpful. In addition, changing the size (length and width) and the material of the patch will change its results. Each design will give a different return loss and different radiation pattern. Not only that, but also the various designs develops distinctive complications that aided in the formation of the final design. Last but not least, the final design is thoroughly discussed and elaborated.

## 11.3. PRIMARY MICRO-STRIP PATCH ANTENNA DESIGN

The micro-strip patch antenna can be constructed on a flat surface, and it consists of three main layers including dielectric substrate, ground plane, and the patch. The antenna is covered with a radiating box (air).

We started with a simple design of a simple primary micro-strip patch antenna using HFSS software. The material for the ground is "Rogers 5880", with a dimension of 100 mm and 90 mm for the X-axis and Y-axis respectively. The substrate has the same dimensions as the ground, while the size of the patch is 40 mm (X-axis) and 30 mm (Y-axis). The operating frequency is the 2.4 GHz, which is the ISM band. The design is shown in the figure below.

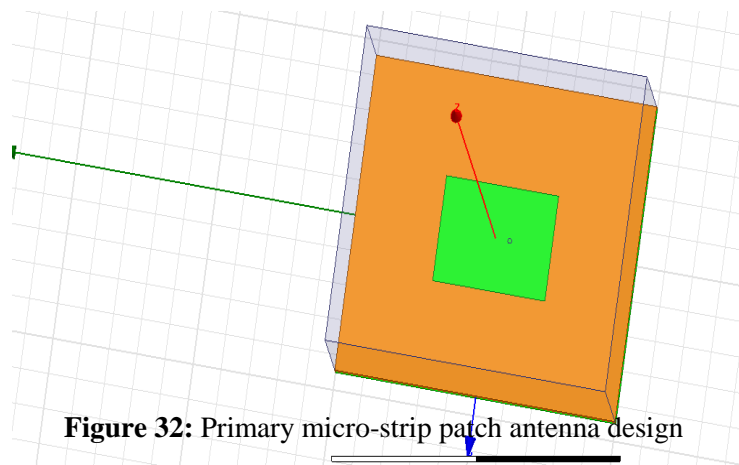

**Figure 32:** Primary micro-strip patch antenna design

Then, the return loss graph was found to determine the reflection coefficient as illustrated in the figure below.

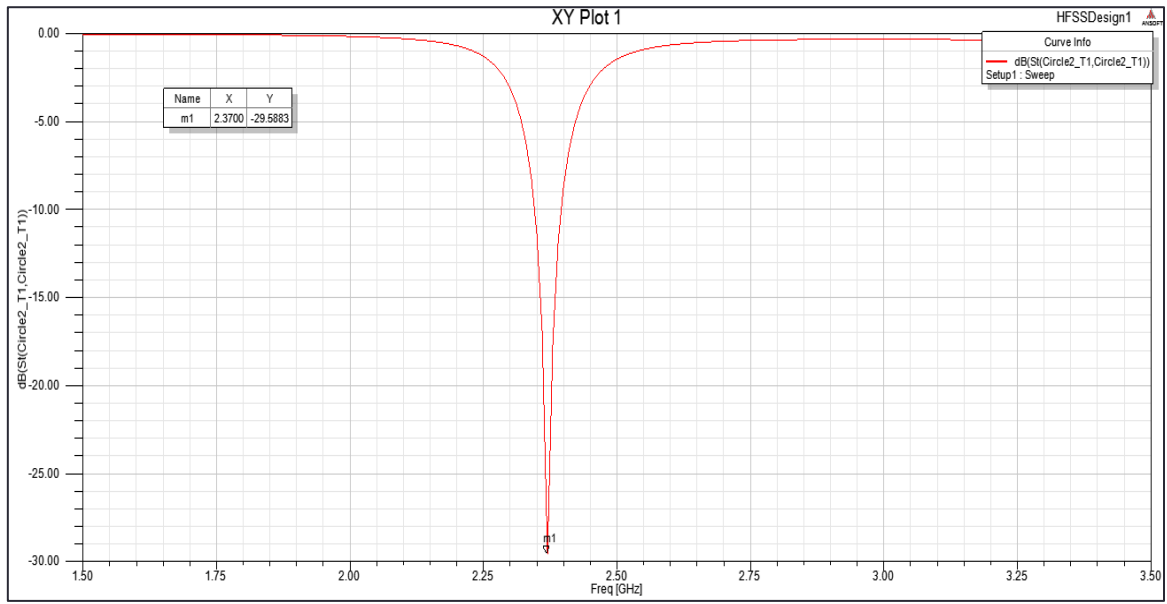

**Figure 33:** Return loss graph for the primary micro-strip patch antenna design

As shown in the above figure, the reflection coefficient ($S_{11}$) at frequency 2.4 GHz is -29.58 dB, which a very good result, as it is below -10 dB.

## 11.4 CIRCULAR MICRO-STRIP PATCH ANTENNA DESIGN

Rather than using a rectangular patch, a circular patch was designed in order to observe the difference in the return-loss parameter. The shape of the designed circular micro-strip patch antenna is illustrated below with several sides in the ANSYS HFSS program. The design's dimensions are same as the previous primary rectangular design, however the radius of the circular patch is 35 mm.

The design contains a dielectric substrate, as a circular radiating patch is on the topside, while a ground plane is in the bottom side. Also, there is an air radiation around the antenna, as shown in Figure 34.

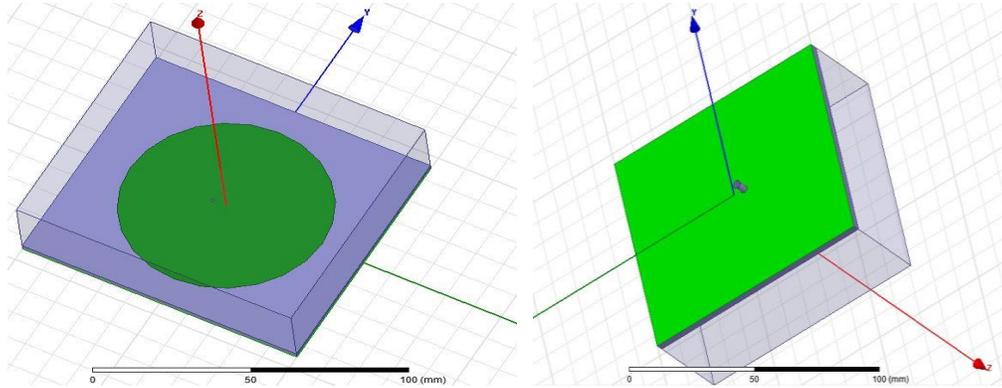

**Figure 34:** Circular Micro-strip Patch Antenna Design

The reflection coefficient |*s*11| of the circular micro-strip patch antenna design is illustrated in the following plot.

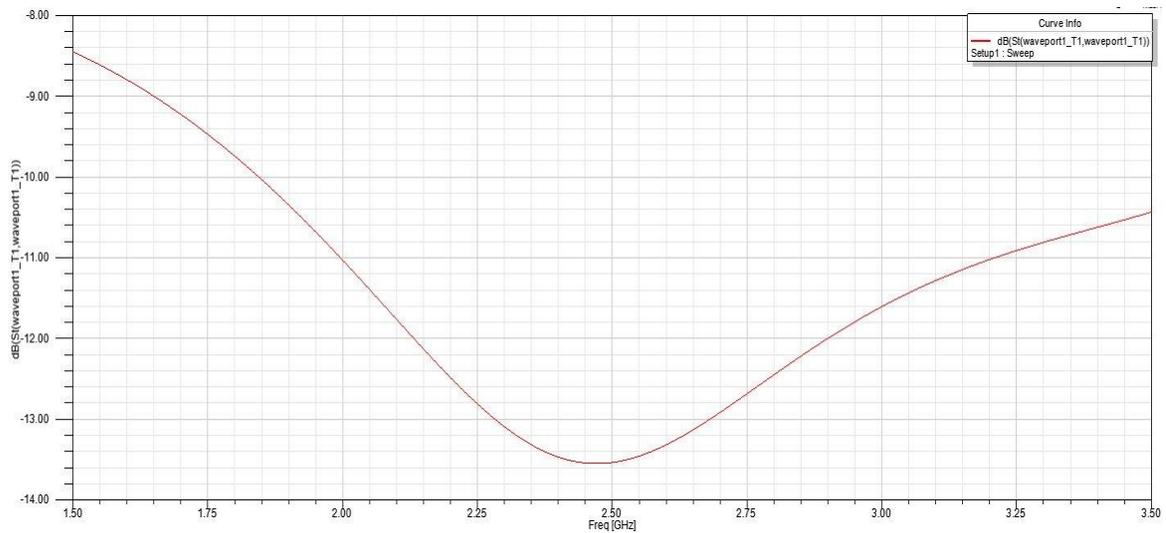

**Figure 35:** Return loss graph for the circular micro-strip patch antenna design

As can be noticed from Figure 35 above, the circular micro-strip patch antenna's reflection coefficient is -13.46 dB over the frequency 2.4 GHz, which means it has higher return-loss than the rectangular micro-strip patch antenna. Thus, the rectangular micro-strip patch antenna is better as it has lower return loss.

## 11.5 SLOTTED RECTANGULAR MICRO-STRIP PATCH ANTENNA DESIGN 1

In this design slotting is applied in order to enhance the return loss. The design consists of three main layers: ground in the bottom, substrate in the middle, and a slotted patch on the top of the substrate. Additively, the antenna is covered with a box of air for radiation. The material of the substrate is copper. The size of the ground was the same as the substrate, which is 37 mm size for X axis and 46mm for Y axis, the patch has 28mm for X-axis and 47mm for Y-axis, while the slot was 12mm for X-axis and 3mm for Y-axis as shown in the figure below. Knowing that the operating frequency was 2.4, which is the ISM band.

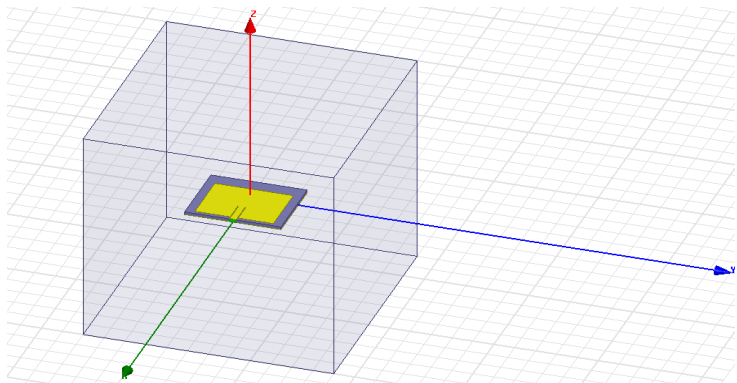

**Figure 36:** Slotted rectangular micro-strip patch antenna design 1

In order to judge the performance of the proposed antenna, radiation pattern in the 3D polar plot and the return loss must be demonstrated.

As it is shown in the figure below, the radiation pattern (3-D polar plot) is crucial, with one side loop, which limit its radiation in the direction of the side loop.

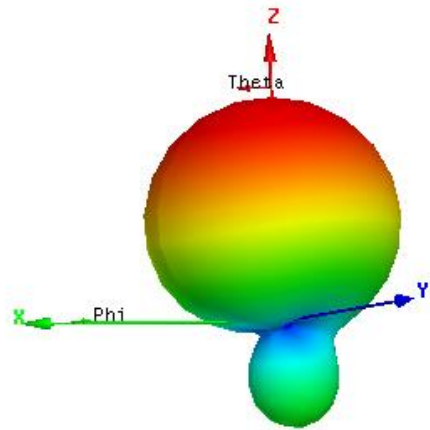

**Figure 37:** 3D polar plot for slotted rectangular micro-strip patch antenna (design 1)

The return loss graph is illustrated in the figure below.

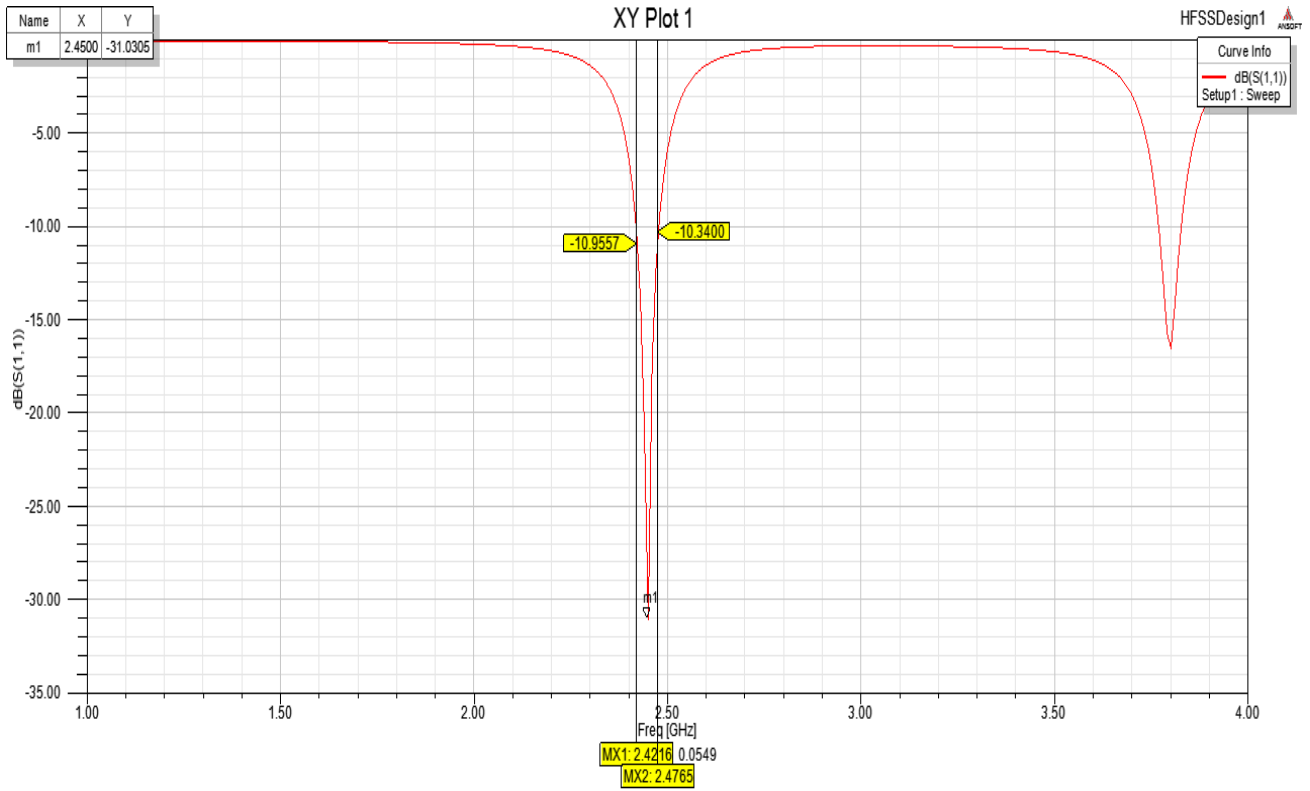

**Figure 38:** Return loss graph for slotted rectangular micro-strip patch antenna (design 1)

From the above figure it is noticeable that at frequency 2.4 GHz the return loss is -33.7, which is much lower than -10dB. Therefore, this design gives a very good and low reflection coefficient, however as the copper is a conductor, which means it has a very high relative permittivity therefore it is not applicable for this medical application.

## 11.6 SLOTTED RECTANGULAR MICRO-STRIP PATCH ANTENNA DESIGN 2

For the purpose of increasing the return loss, a slotted rectangular micro-strip patch antenna was designed. The antenna operates in the ISM band at 2.45 GHz. This design consist of a ground in the bottom , middle slot in the patch which is on the top of the substrate, a feed line port and air radiation. The material of the substrate is FR4_epoxy which has a relative permittivity equal to 4.4. The slot is in the middle of the patch with 20mm size for both X and Y axis, while the patch has 50mm for X and Y axis, as shown in Figure 39.

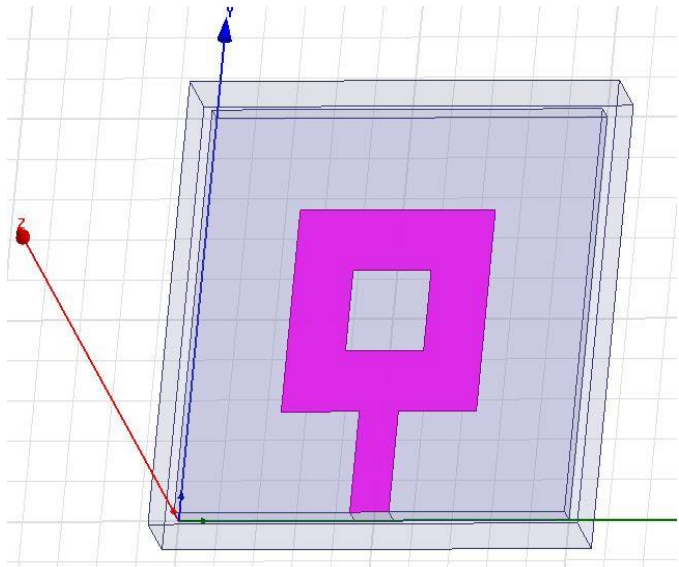

**Figure 39**: Slotted rectangular micro-strip patch antenna (middle slot)

As the antenna design performance is measured through the operating resonance frequency the return loss parameter, both the radiation pattern in the 3D polar plot and the return loss will be illustrated below with a frequency solution equal to 2.45 GHz. The radiation pattern is crucial since it demonstrates the communication with the receiver from all directions.

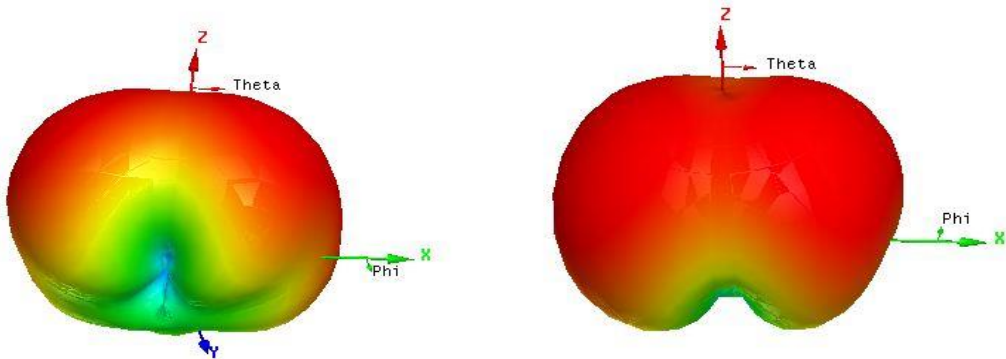

**Figure 40:** 3D polar plot for slotted rectangular micro-strip patch antenna (middle slot)

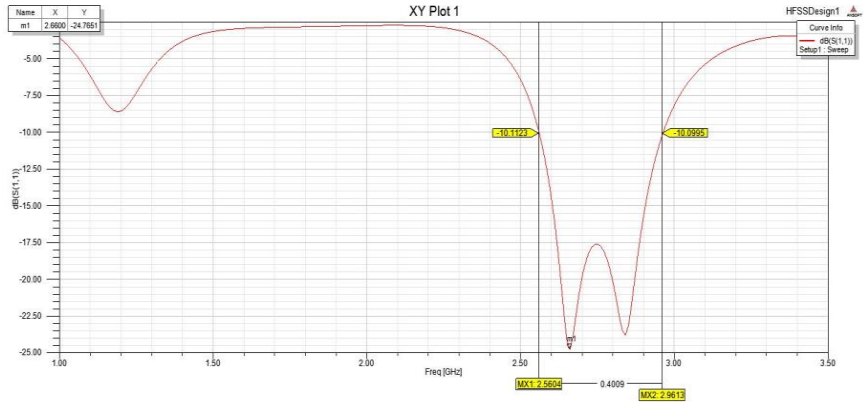

**Figure 41:** The return loss for slotted rectangular micro-strip patch antenna (middle slot)

As can be noticed from Figure 41, the bandwidth is 0.4009 GHz at -10 dB and the return loss is less than -10 dB which is -24.5 dB at a frequency of 2.65 GHz which means that the frequency is shifted and it is not in ISM band. Therefore, some changes must be done in order to have a return loss that is less than -10 dB at a frequency equals to the ISM frequency range from 2.4 to 2.5 GHz.

## 11.7 SLOTTED RECTANGULAR MICRO-STRIP PATCH ANTENNA <u>FINAL DESIGN</u>

The first change was related to the material of the substrate, where the FR4_epoxy material was changed to Rogers RT/duroid 5880 (tm) material, which has a relative permittivity equal to 2.2. Moreover, the size of the patch was changed to 35mm for the X-axis and 40mm for the Y axis, while keeping the slot's size fixed, but its location was changed, as shown in the figure below.

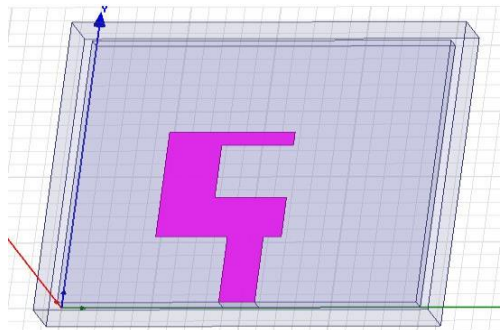

**Figure 42:** Slotted rectangular micro-strip patch antenna design (side slot)

The radiation pattern in the 3D polar plot is illustrated below with a frequency solution equal to 2.45 GHz.

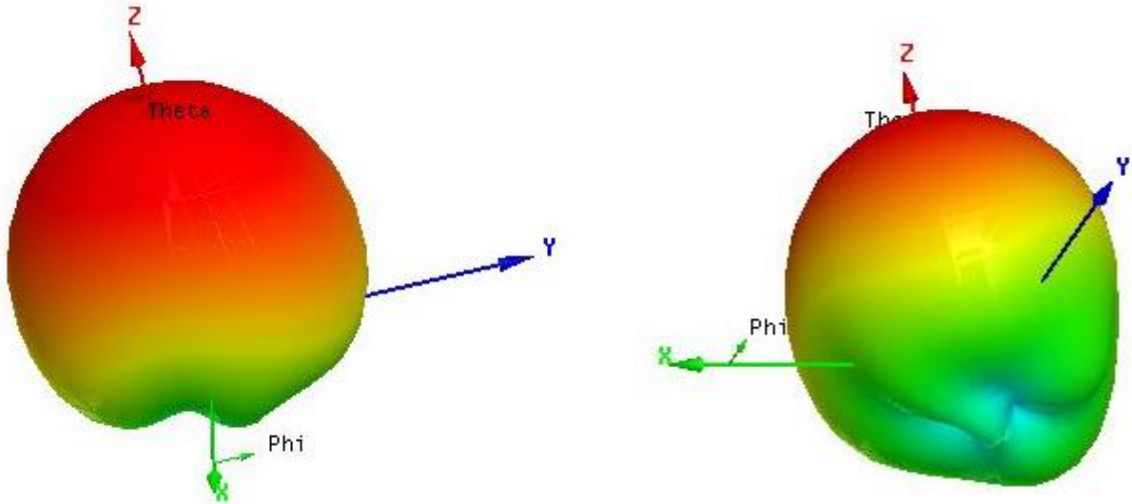

**Figure 43:** 3D polar plot for slotted rectangular micro-strip patch antenna (side slot)

Finally, the return loss graph was found, as shown below.

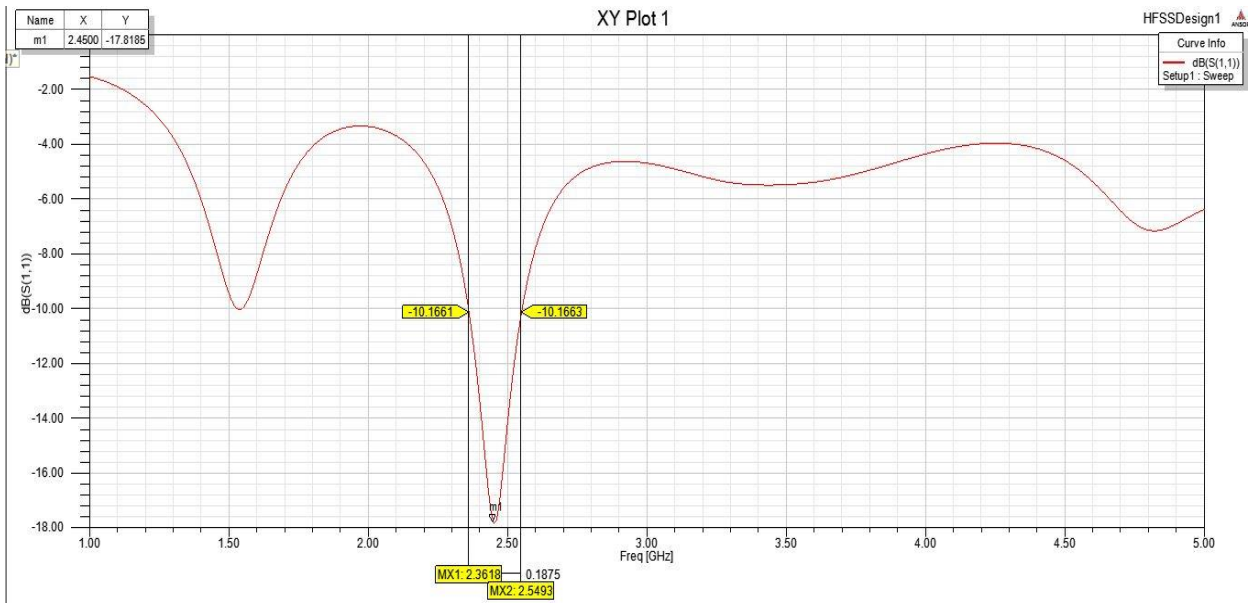

**Figure 44:** The return loss for slotted rectangular micro-strip patch antenna (side slot)

It can be noticed from Figure 44 that the bandwidth is 0.1875 GHz at -10 dB, which is lower than the bandwidth for the previous middle slot design. In addition, the return loss is equal to -17.8 dB which is less than -10 dB, and it is lower than the previous middle slot design, however, it operates in ISM band at a frequency of 2.45 GHz.

## 11.8. COMPARISON BETWEEN ALL THE DESIGNS

The table below illustrates a comparison between the five proposed designs according to the return loss, -10 dB bandwidth, the dimension of the patch, and finally the limitations of each design.

**Table 3:** comparison between the five designs

| | Return Loss $S_{11}$ | -10dB Bandwidth | Dimension of the Patch | Limitations |
|---|---|---|---|---|
| **Primary design** | -29.58 dB | 2.34 to 2.4 GHz | Width 30 mm Length 40 mm | |
| **Circular design** | -13.46 dB | 1.843 to 3.703 GHz | Radius 35 mm | The return loss is higher than the rectangular patch. |
| **Slotted Design 1** | -33.7 dB | 2.42 to 2.47 GHz | Width 28.6 mm Length 37.5 mm | The patch material (copper) is conductor |
| **Slotted Design 2** | -24.76 dB | 2.56 to 2.96 GHz | Width 50 mm Length 50 mm | The frequency is shifted (not at 2.4 GHz) |
| **Slotted Final Design** | -17.8 dB | 2.36 to 2.54 GHz | Width 35 mm Length 40 mm | |

## 11.9. REMARKS

In order to reach to the final design, many designs were done and repeated again and again by changing the material and dimension in order to enhance the return loss. From the two primary designs (rectangular and circular) we identified that for this application using a rectangular patch antenna will give better results. In addition to that, applying slotting to the patch resulted in a greater design. However, changing the size of slot and/or its position caused a change on the band of the -10dB bandwidth. Finally, Each of the proposed designs had its own limitation, and we chose the most suitable one for this medical application.

# CHAPTER TWELVE

# **CONCLUSIONS, FUTURE DIRECTIONS, AND CRITICAL APPRAISAL**

**12.1.CONCLUSIONS**

This project demonstrated several antenna types and designs in order to have a significant understanding about designing a compact and more efficient Micro-Strip Patch Antenna design to be used in medical and breast cancer diagnosis. This included the understanding of the Micro-Strip Patch Antenna characteristics, applications, and its operational design. Moreover, the use of HFSS software and being familiar with its different functions was very crucial to be able to design a biomedical, compatible, and efficient antenna that can be utilized for breast cancer detection purposes.

The Discussion of the main qualities, characteristics, and advantages of Micro-Strip Patch Antenna was highlighted in order to illustrate its importance and the reason for choosing it for this project. It has a high gain and directivity, low profile, easy to construct, and efficient return loss. Then, the challenges faced through designing wearable antenna were discussed, including battery life, size limitation, and reduced antenna efficiency. This reduction is mainly because of the transformation of the electric fields into heat by the human body, however, it can be overcome by a suitable impedance matching, as the transmitted power approximately equals to the reflected power. Regard to the size limitation, this challenge can be solved by some methods, including shorting pin, slotting, and stacking. In addition, minimizing the physical size, reflection coefficient reduction, electrical size increment, omnidirectional radiation pattern, biocompatibility, scattering parameter, and standing wave ratio were discussed in details. Furthermore, the ISM band operating frequency was used in this project, which is dedicated for medical wearable antennas. After that, the main features and advantages of the High Frequency Structure Simulator (HFSS) software was mentioned which was utilized to achieve the goals of our project.

After that, the specifications and basics of Micro-Strip Patch Antenna design were highlighted as well as a new method for return loss enhancement, which is Meta-Materials (MTMs). Finally, the discussion of various antenna designs and their results, including the final design were mentioned, each design has its own advantages and drawbacks, which were obtained in this project by using the HFSS software. These several designs were developed in order to indicate the major obstacles that were faced until we reach the final design. It was also noticed that in order to enhance the return loss, the resonance frequency might be shifted to undesirable ones.

At the end, we can conclude that this project has developed our skills in facing and solving obstacles in real-life-engineering. In addition to, all team members have benefit from this project, since breast cancer detection applications are very significant and extremely sensitive, as a lot of effort was required to learn about the combination of engineering and medical fields.

## 12.2. RECOMMENDATIONS FOR FUTURE DIRECTIONS

After finishing this report, it is significant to mention and recommend some future directions of this project in order to get a design with more efficiency taking into consideration the size of the antenna, return-loss, and radiation pattern.Firstly, as the device will be wearable, the size is important and the smaller the antenna the better. However, it must be done while taking into consideration the high performance (increase it or at least keep it as it is). Thus, even though having a compact antenna is needed the performance of the antenna is also significant. So, this task can be considered as one of the most challenging tasks. Secondly, the reflection coefficient must be considered as well. It is challenging to reduce the return loss without increasing the relative permittivity and the dimension of the patch, however, the smaller the return loss the better.

## 12.3. CRITICAL APPRAISAL

This project is considered as a very challenging project, considering the various topics included to reach to the final design. Different engineering fields were involved in this project, for instance: communication engineering, electronic engineering and biomedical, in addition to the instrumentation and electromagnetics. Furthermore, in this project a new software was used the High Frequency Structure Simulator, which is different from the other software programs that we learned about during our academic years in the university such as MATLAB and C++. HFSS was used for its high accuracy and reliable results, however it took as a lot of time to learn how to use this software correctly, as we watched a lot of tutorials on YouTube, in addition to reading some papers and books about it.

It took us two semesters (Spring 2016 and Fall 2016) to complete this project. It was challenging to meet the deadlines as all the deadlines were close to each other, considering all the other academic courses we are taking. However, having a good team with an organized team leader and a motivate project supervisor who always supported us helped us in finishing this project on time.

Finally, during this project we have gained advanced knowledge about biomedical, antenna and healthcare fields. In addition, it was nice how we were able to apply what we learned in communication engineering in another field such as the medical field to help in saving others lives.

## III. References:

[1] R. M. Shubair, "Robust adaptive beamforming using LMS algorithm with SMI initialization," in 2005 IEEE Antennas and Propagation Society International Symposium, vol. 4A, Jul. 2005, pp. 2–5 vol. 4A.

[2] R. M. Shubair and W. Jessmi, "Performance analysis of SMI adaptive beamforming arrays for smart antenna systems," in 2005 IEEE Antennas and Propagation Society International Symposium, vol. 1B, 2005, pp. 311–314 vol. 1B.

[3] F. A. Belhoul, R. M. Shubair, and M. E. Ai-Mualla, "Modelling and performance analysis of DOA estimation in adaptive signal processing arrays," in 10th IEEE International Conference on Electronics, Circuits and Systems, 2003. ICECS 2003. Proceedings of the 2003, vol. 1, Dec. 2003, pp. 340–343 Vol.1. 19

[4] R. M. Shubair and A. Al-Merri, "Robust algorithms for direction finding and adaptive beamforming: performance and optimization," in The 2004 47th Midwest Symposium on Circuits and Systems, 2004. MWSCAS '04, vol. 2, Jul. 2004, pp. II–589–II–592 vol.2.

[5] E. Al-Ardi, R. Shubair, and M. Al-Mualla, "Direction of arrival estimation in a multipath environment: An overview and a new contribution," in ACES, vol. 21, 2006.

[6] G. Nwalozie, V. Okorogu, S. Maduadichie, and A. Adenola, "A simple comparative evaluation of adaptive beam forming algorithms," International Journal of Engineering and Innovative Technology (IJEIT), vol. 2, no. 7, 2013.

[7] M. A. Al-Nuaimi, R. M. Shubair, and K. O. Al-Midfa, "Direction of arrival estimation in wireless mobile communications using minimum variance distortionless response," in Second International Conference on Innovations in Information Technology (IIT'05), 2005, pp. 1–5.

[8] M. Bakhar and D. P. Hunagund, "Eigen structure based direction of arrival estimation algorithms for smart antenna systems," IJCSNS International Journal of Computer Science and Network Security, vol. 9, no. 11, pp. 96–100, 2009.

[9] M. AlHajri, A. Goian, M. Darweesh, R. AlMemari, R. Shubair, L.Weruaga, and A. AlTunaiji, "Accurate and robust localization techniques for wireless sensor networks," June 2018, arXiv:1806.05765 [eess.SP].

[10] J. Samhan, R. Shubair, and M. Al-Qutayri, "Design and implementation of an adaptive smart antenna system," in Innovations in Information Technology, 2006, 2006, pp. 1–4.

[11] M. AlHajri, A. Goian, M. Darweesh, R. AlMemari, R. Shubair, L.Weruaga, and A. Kulaib, "Hybrid rss-doa technique for enhanced wsn localization in a correlated environment," in Information and Communication Technology Research (ICTRC), 2015 International Conference on, 2015, pp. 238–241.

[12] L. Mohjazi, M. Al-Qutayri, H. Barada, K. Poon, and R. Shubair, "Deployment challenges of femtocells in future indoor wireless networks," in GCC Conference and Exhibition (GCC), 2011 IEEE. IEEE, 2011, pp. 405–408.

[13] C. A. Balanis, "Antenna Theory Analysis and Design", Third Edition 2005.

[14] R. M. Shubair and H. Elayan, "In vivo wireless body communications: State-of-the-art and future directions," in Antennas & Propagation Conference (LAPC), 2015 Loughborough. IEEE, 2015, pp. 1–5.

[15] H. Elayan, R. M. Shubair, J. M. Jornet, and P. Johari, "Terahertz channel model and link budget analysis for intrabody nanoscale communication," IEEE transactions on nanobioscience, vol. 16, no. 6, pp. 491–503, 2017.

[16] H. Elayan, R. M. Shubair, and A. Kiourti, "Wireless sensors for medical applications: Current status and future challenges," in Antennas and Propagation (EUCAP), 2017 11th European Conference on. IEEE, 2017, pp. 2478–2482.

[17] H. Elayan and R. M. Shubair, "On channel characterization in human body communication for medical monitoring systems," in Antenna Technology and Applied Electromagnetics (ANTEM), 2016 17th International Symposium on. IEEE, 2016, pp. 1–2.

[18] H. Elayan, R. M. Shubair, A. Alomainy, and K. Yang, "In-vivo terahertz em channel characterization for nano-communications in wbans," in Antennas and Propagation (APSURSI), 2016 IEEE International Symposium on. IEEE, 2016, pp. 979–980.

[19] H. Elayan, R. M. Shubair, and J. M. Jornet, "Bio-electromagnetic thz propagation modeling for in-vivo wireless nanosensor networks," in Antennas and Propagation (EUCAP), 2017 11th European Conference on. IEEE, 2017, pp. 426–430. 21

[20] H. Elayan, C. Stefanini, R. M. Shubair, and J. M. Jornet, "End-to-end noise model for intra-body terahertz nanoscale communication," IEEE Transactions on NanoBioscience, 2018.

[21] H. Elayan, P. Johari, R. M. Shubair, and J. M. Jornet, "Photothermal modeling and analysis of intrabody terahertz nanoscale communication," IEEE transactions on nanobioscience, vol. 16, no. 8, pp. 755–763, 2017.

[22] H. Elayan, R. M. Shubair, J. M. Jornet, and R. Mittra, "Multi-layer intrabody terahertz wave propagation model for nanobiosensing applications," Nano Communication Networks, vol. 14, pp. 9–15, 2017.

[23] H. Elayan, R. M. Shubair, and N. Almoosa, "In vivo communication in wireless body area networks," in Information Innovation Technology in Smart Cities. Springer, 2018, pp. 273–287.

[24] M. O. AlNabooda, R. M. Shubair, N. R. Rishani, and G. Aldabbagh, "Terahertz spectroscopy and imaging for the detection and identification of illicit drugs," in Sensors - Networks Smart and Emerging Technologies (SENSET), 2017, 2017, pp. 1–4.

[25] M. Nalam, N. Rani and A. Mohan, "Biomedical Application of Microstrip Patch Antenna," International Journal of Innovative Science and Modern Engineering (IJISME), Vol. 2, Issue 6, May 2014.

[26] S. Mishra and M. Bhat. Microstrip Patch Antenna [Online], Available: http://conference.aimt.edu.in/paper/ec%20paper/Microstrip%20Patch%20Antenna.pdf

[27] Fear, E. C., P. M. Meaney, and M. A. Stuchly, "Microwaves for breast cancer detection," IEEE Potentials, 12–18, Feb./Mar. 2003.

[28] Y .Rahmat-Samii and J.Kim," Implanted Antennas in Medical Wireless Communications," A Publication in the Morgan and Claypool Publishers' series, 1st edition, vol.01, 2006.

[29] "Challenges in the Design of Microwave Imaging Systems for Breast Cancer Detection", Technical University of Denmark, Volume 11, Number 1, 2011.

[30] K. Koski, E. Moradi, T. Björninen, L. Sydänheimo, Y. Rahmat-Samii, and L. Ukkonen, "On-Body Antennas: Towards Wearable Intelligence," IEEE Conference Publications, 2014.

[31] K. Agarwal, Yong-Xin Guo, "Interaction of Electromagnetic Waves with Humans in Wearable and Biomedical Implant Antennas," IEEE Conference Publications, pp. 154- 157, 2015.

[32] G. Marin, G. Samoilescu, A. Baciu, D. Iorgulescu, and S. Radu, "Assessment of the Need for Protection against Electromagnetic Radiation of Personnel Onboard Warships," IFMBE Proceedings, Vol. 44, pp. 285-290, 2014.

[33] P. J. Bevelacqua. Introduction to Impedance Matching [Online]. Available: http://www.antenna-theory.com/tutorial/smith/smithchart5.php

[34] Integrated Publishing, Inc. Basic Antennas [Online]. Available: http://electriciantraining.tpub.com/14182/css/14182_184.htm

[35] W.S. Kang, J. A. Park and Y. J. Loon, "Simple Reconfigurable Antenna with Radiation Pattern," Electronics Letters, Vol. 44, Issue 3, pp. 182-183, January 2008.

[36] A. Kiourti and K. S. Nikita, "A Review of Implantable Patch Antennas for Biomedical Telemetry: Challenges and Solutions," IEEE Antennas and Propagation Magazine, Vol. 54, Issue 3, pp. 211-214, June 2012.

[37] Y. Rahmat-Samii , "Wearable and Implantable Antennas: Applications in Communications, Monitoring and Diagnostics," IEEE Conference Publications, 2014.

[38] "S-Parameters," for Antennas (S11, S12, ...), 2015. [Online]. Available: http://www.antenna-theory.com/definitions/sparameters.php. [Accessed: 22-Apr-2016].

[39] daeein walraven, "VSWR," VSWR, 2015. [Online]. Available: http://www.antenna-theory.com/definitions/vswr.php. [Accessed: 27-Jun-2016].

[40] Wasusathien, Wittawat et al. "Ultra Wideband Breast Cancer Detection By Using SAR For Indication The Tumor Location". Waset.org. N.p., 2014. Web. 8 June 2016.

[41] R. Brinda, and S. Preethy, "Miniaturized Antenna with Combination of Meander and Square Spiral Slots for Biomedical Applications," International Journal of Computer Applications, Vol. 85, No. 4, January 2014.

[42] "// Encyclopedia," ISM band Definition from PC Magazine Encyclopedia. [Online]. Available: http://www.pcmag.com/encyclopedia/term/45467/ism-band.

[43] J. P. Stang, "A tapered microstrip patch antenna array for use in breast cancer screening via 3D active microwave imaging," IEEE Xplore Document - A tapered microstrip patch antenna array for use in breast cancer screening via 3D active microwave imaging, 2009. [Online]. Available: http://ieeexplore.ieee.org/document/5171907/. [Accessed: 15-May-2016].

[44] M. Klemm, I.J. Craddock, J. Leendertz, A.W. Preece, R. Benjamin, "BREAST CANCER DETECTION USING SYMMETRICAL ANTENNA ARRAY", University of Bristol, Bristol, UK.

[45] Hamed M. Jafari, M. Jamal Deen, Steve Hranilovic, and Natalia K. Nikolova, "A Study of Ultrawideband Antennas for Near-Field Imaging", IEEE TRANSACTIONS ON ANTENNAS AND PROPAGATION, VOL. 55, NO. 4, APRIL 2007.

[46] "ANSYS HFSS for Antenna Simulation," ANSYS, 2014. [Online]. Available: http://resource.ansys.com/staticassets/ansys/staticassets/resourcelibrary/techbrief/ab-ansys-hfss-for-antenna-simulation.pdf.

[47] Latest HFSS Release Improves Speed and Accuracy. (n.d.). Retrieved from www.microwavejournal.com: http://www.microwavejournal.com/articles/2029-latest-hfss-release-improves-speed-and-accuracy

[48] "Microstrip (Patch) Antennas," Microstrip Antennas: The Patch Antenna, 2011. [Online]. Available: http://www.antenna-theory.com/antennas/patches/antenna.php.

[49] T. M., "MICROSTRIP PATCH ANTENNAS FED BY SUBSTRATE INTEGRATED WAVEGUIDE," 2013. [Online]. Available: https://www.vutbr.cz/www_base/zav_prace_soubor_verejne.php?file_id=70002.

[50] S. Mishra and M. Bhat. Microstrip Patch Antenna [Online], Available: http://conference.aimt.edu.in/paper/ec%20paper/Microstrip%20Patch%20Antenna.pdf

[51] Fear, E. C., P. M. Meaney, and M. A. Stuchly, "Microwaves for breast cancer detection," IEEE Potentials, 12–18, Feb./Mar. 2003.

[52] P. M. Kannan and V. Palanisamy, "Dual Band Rectangular Patch Wearable Antenna on Jeans Material," Engineering and Technology, 2012.

[53] Dual Band Rectangular Patch Wearable Antenna on Jeans MaterialP.Muthu Kannan, V.Palanisamy, Department of ECE, Associate Prof. & Head i/c, Department of CSE, Galgotias University, Alagappa University, Greater Noida, India. Karaikudi, India.

[54] "multiband antenna," What is multiband antenna? Webopedia Definition. [Online]. Available: http://www.webopedia.com/TERM/M/multiband_antenna.html.

[55] S. M. Aguilar, M. A. Al-Joumayly, M. J. Burfeindt, N. Behdad, and S. C. Hagness, "Multi-Band Miniaturized Patch Antennas for a Compact, Shielded Microwave Breast Imaging Array," IEEE transactions on antennas and propagation, 2013. [Online]. Available: https://www.ncbi.nlm.nih.gov/pmc/articles/PMC4226417/

[56] T. Cai, G. Wang, X. Zhang, Y. Wang, B. Zong, and H. Xu, "Compact Microstrip Antenna With Enhanced Bandwidth by Loading Magneto-Electro-Dielectric Planar Waveguided Metamaterials," IEEE Transaction on Antennas and Propagation, Vol. 63, Issue 5, May 2015, pp. 2306-2311.

[57] M. S. Khan, A. D. Capobianco, S. M. Asif, D. E. Anagnostou, R. M. Shubair, and B. D. Braaten, "A Compact CSRR-Enabled UWB Diversity Antenna," IEEE Antennas and Wireless Propagation Letters, vol. 16, pp. 808–812, 2017.

[58] R. M. Shubair and Y. L. Chow, "A closed-form solution of vertical dipole antennas above a dielectric half-space," IEEE Transactions on Antennas and Propagation, vol. 41, no. 12, pp. 1737–1741, Dec. 1993.

[59] A. Omar and R. Shubair, "UWB coplanar waveguide-fed-coplanar strips spiral antenna," in 2016 10th European Conference on Antennas and Propagation (EuCAP), Apr. 2016, pp. 1–2.

[60] Institute of Physics. Metamaterials [Online], Available: http://www.iop.org/resources/topic/archive/metamaterials/